\documentclass{elsart}
\usepackage{graphicx}
\def\contrac#1#2#3<#4>{%
\setbox1=\hbox{$#1$}%
\setbox2=\hbox{$#2$}%  
\setbox3=\hbox{$#3$}%
\dimen0=.5\wd1%
\advance\dimen0 by \wd2%
\advance\dimen0 by .5\wd3%
\setbox0=\hbox{\kern.5\wd1%
\vtop{\hbox to \dimen0{\vrule depth#4\hfil\vrule}\hrule}}%
\dimen0=#4 \advance\dimen0 by \lineskip \dp0=\dimen0%
\wd0=0pt%
\setbox1=\vtop{\lineskiplimit=10cm\lineskip=1mm \box1\box0}%
\box1\box2\box3}
\newcommand{\ide}{\mbox{\Large\bf $1\!\!1$}}
\newcommand{\img}{{\mathrm{i}}}
\begin{document}
\begin{frontmatter}
\title{\Large \bf Particle number projection with effective
forces\thanksref{ring}.}
\author{M. Anguiano\thanksref{marta}},
\author{J.L. Egido},
\author{L.M. Robledo},  
\address{Departamento de F\'{\i}sica
Te\'orica, Universidad Aut\'onoma de Madrid, E-28049 Madrid, Spain} 
\date{\today}
\thanks[ring]{Dedicated to Prof. Peter Ring on the occasion of his
60$^{\rm{th}}$ birthday.}
\thanks[marta]{Present address: Dipartamento di Fisica, Universita di Lecce,
73100 Lecce, Italy.}
\begin{abstract}
The particle number projection method is formulated for density dependent
forces and in particular for  the finite range Gogny force.  
Detailed formula for the  projected energy and its gradient are provided.
 The problems arising from the neglection  of any exchange term, which may 
lead to divergences, are thoroughly discussed and the possible inaccuracies
estimated. Numerical results for the projection after variation method are
shown for the nucleus $^{164}$Er  and for the projection before variation
approach for the nuclei $^{48,50}$Cr. We also confirm the Coulomb antipairing 
effect found in mean field theories. 

\end{abstract}
\end{frontmatter}

%
%\newpage
%
\section{Introduction}
  The simplest approach to the many-body system is the mean--field approximation (MFA). 
The most powerful MFA is the Hartree--Fock--Bogoliubov (HFB) theory, where
particle-hole  and particle-particle correlations are taken into account at the same foot. 
The success of this simple approach is based on the large variational Hilbert space
generated by  symmetry breaking wave functions. For an atomic nucleus and in the most
 general HFB approach as many symmetries
as possible, in particular the rotational invariance and the particle number
symmetries are broken. In these approaches the quantum numbers are only
conserved on the average, a semiclassical approach to the full quantum requirement to
the wave function of being an eigenstate of the symmetry operators. It is well known
\cite{RS.80} that the plain HFB approach works well in the case of a strongly correlated regime
and with a large number of nucleons participating in the symmetry breaking process.
In the case of the rotational invariance, the ideal 
situation is  represented  by a strongly deformed 
heavy nucleus where a large number of nucleons contribute to the collective phenomenon
of deformation.  In the case  of the particle number symmetry, in which we are interested 
in this paper, the situation is far from being 
satisfactory because only a few pairs of nucleons
( four to five)  in the vicinity of the Fermi surface are thought to cause the phenomenon 
of nuclear superfluidity. Thus, it is well known that the treatment of  pairing 
correlations in the mean field approach produces an unrealistic transition from the 
superfluid to the normal phase along the yrast band, not expected in a finite system. 
In order to get an improved description of the nucleus it is clear that one has to
go beyond the mean field approximation to include correlations. This task can
be achieved in a simple and effective way within the 
particle number projection (PNP) \cite{PNP.orig} formalism,
another possibility is to build correlations by the Random-Phase-Approximation (RPA)
\cite{EMR.80}. The latter is, however,  from the applications point
of view, very limited, since the required calculation of the ground state correlations
can be performed only for very simple models and/or separable interactions. 
 Concerning this point one has to keep in mind that  not only the RPA but 
 also other
sophisticated theories  have been developed only in simple models or with separable
forces. Thus the first  calculation with  exact particle number projection before 
the variation  was done about twenty years ago \cite{Egi82} with the pairing 
plus quadrupole Hamiltonian and the Baranger-Kumar configuration space.

In the  last years in an attempt to achieve some progress in calculations with
effective forces and  large configuration spaces, the  Lipkin Nogami method 
\cite{LN.orig} has experienced a revival. The  great advantage of the 
Lipkin Nogami method is that corrects, in an approximate way, for the zero point
energy associated with the breakdown of the particle number symmetry at the 
{\em mean field level}, i.e. the calculations do not get much involved as
compared to the full HFB ones. Some applications of this method with effective
forces have been done  with the Skyrme force \cite{Gall94,Bon96},
the Gogny force\cite{Val96} and the relativistic mean field approach \cite{Ring.LN}. 
The Lipkin Nogami method on the other hand is  just a recipe
and cannot be obtained from a variational principle. It is therefore desirable to
develop an exact particle number projection which can be applied to effective
forces.
 The purpose of this paper is to develop an exact particle
number projection before (and after) the variation  for density dependent
forces and in particular for the Gogny force. 
As we shall see, the exchange terms play an important role in  PNP,
in fact the neglection of any exchange term may lead to divergences.
 As it is well known, in general, a zero range force is
unable to provide good pairing properties. As a matter of fact, normally
the pairing terms of the force are neglected in order to get reasonable 
Hartree Fock results. To perform HFB calculations a pairing force has been 
usually added ad hoc.
The finite range of the density dependent Gogny force, on the other hand, 
provides good pairing properties without neglection of any exchange term.
In fact, the force was designed with this
intention\cite{Gog80} and  this property makes
the  Gogny force one of the few  effective density dependent forces for
 which particle number projection is feasible without  problem due to
 divergences. 
All the Skyrme parametrizations (with the
few exceptions of those which provide reasonable pairing correlations)
 and all relativistic mean field approaches  may  give in general
divergent contributions to the projected energy. 

 In section~\ref{sec:GPT}, we introduce 
the particle number projection method for non-separable density dependent 
forces. In section~\ref{sec:DDP} we present the formalism for the density
dependent part of the interaction and discuss some alternative prescriptions
to this dependence.
The gradient of the  projected energy  to be implemented in the calculations 
is developed in section~\ref{sec:VAP}. 
In section~\ref{sec:convL}, we present results for the projection after variation 
(PAV) method  for 
the nucleus $^{164}$Er and discuss the problems associated with the divergent
terms. Results for the variation after projection (VAP) method for the
nuclei $^{48,50}$Cr  are discussed in section~\ref{sec:VAPR}.

\section{Particle Number Projected Hartree--Fock--Bogoliubov Theory.}
\label{sec:GPT}

  The Particle Number Projected Hartree--Fock--Bogoliubov theory 
  is the simplest symmetry conserving mean field approach
which takes into account particle-particle correlations. This theory has been
derived in the past for non-density dependent interactions and a detailed
formulation has  only been given for separable forces like the Pairing
Plus Quadrupole model \cite{Egi82}.
 In order to generalize this theory to non-separable and
  density dependent forces, like the Gogny 
or the Skyrme force, and to introduce the pertinent notation we shall also 
summarize the known formulation.

Let $|\Phi \rangle$  be a product wave function of the HFB type, i.e, the vacuum 
to the quasiparticle operators given by the Bogoliubov transformation 
%
%%%%%%%%%%%%%%BOGOLIUBOV TRANSFORMATION%%%%%%%%%%%%%%%%%%%
%
\begin{equation}
{\alpha}_{\mu} = \sum_k U^*_{k\mu}c_k + V^*_{k\mu}c_k^{\dagger},
\label{BogT}
\end{equation}
where $(c_k, c_k^{\dagger})$ are the particle annihilation and creation 
operators (usually in the  Harmonic Oscillator basis). 
This transformation does not conserve the particle number symmetry and the wave
function $|\Phi \rangle$  is not an eigenstate of the particle number 
operator $\hat{N}$. One can generate, however,
out of  $|\Phi \rangle$,  an eigenstate, $|\Psi_N \rangle$, of  $\hat{N}$
by the well known projection technique \cite{RS.80}
%
%%%%%%%%%%%%%%%%------PROJECTED WAVE FUNCTION------%%%%%%%%%%%%%%%%%% 
%
\begin{equation}
|{\Psi}_N \rangle = {\hat{P}}^N |\Phi \rangle = \frac{1}{2\pi}\int_{0}^{2\pi} 
d{\varphi}\,e^{i \varphi (\hat{N}-N)} \, |\Phi \rangle,
\label{PWF}
\end{equation} 
where the integration interval can be reduced to $[0,\pi]$ whenever
the intrinsic wave function has a well defined ``number parity'' \cite{RS.80}
quantum number.  Furthermore,
 the integral can be discretized in a sum using the Fomenko expansion \cite{Fom}
%
%%%%%%%%%%%%%%%%-------FOMENKO EXPANSION------------%%%%%%%%%%%%%%%%%%%%        
%
\begin{equation}
{{\hat{P}}_L}^N = \frac{1}{L} \sum_{l=1}^L e^{i{\varphi}_l(\hat{N}-N)} , 
\, \,\, \,\, \,\, \,
{\varphi}_l=\frac{\pi}{L}l,
\label{Fom-exp}
\end{equation}
where $L$ is the number of points used in the calculations, for $L\rightarrow \infty$
one gets the exact solution, usually $L=8$ is used in the calculations.
$P^N$ projects only on one system of particles,
in the general case, however, one has to perform a double projection, for the number
of neutrons and protons. In this case  the projected matrix element of any operator 
$\hat{O}$, such that  $[\hat{O},\hat{N}]=0$, is given by \cite{Egi82}
%
%%%%%%%%%%%%%%%%%---------PROJECTED MATRIX ELEMENT---%%%%%%%%%%%%%%%%%%%%%
%
%
\begin{equation}
O^P  = \frac{\langle \Phi| \hat{O} \hat{P}^{N_{\pi}} \hat{P}^{N_{\nu}} | \Phi
\rangle } { \langle \Phi |\hat{P}^{N_{\pi}} \hat{P}^{N_{\nu}} \ \Phi \rangle} =
 \sum_{l_{\pi}=1}^L \sum_{l_{\nu}=1}^L
{\displaystyle y_{l_{\pi}} y_{l_{\nu}} } O({\varphi}_{\pi},{\varphi}_{\nu}), 
\label{PME}
\end{equation}
where we have introduced
%
%%%%%%%%%%-----------DEFINITION OF y_l--------%%%%%%%%%%%%%%%%%%%%%%
\begin{equation}
{\displaystyle y_{l_{\tau}} }  =  \frac{\langle {\Phi}_{\tau} | 
 e^ {i {\varphi}_{l_{\tau}} ({\hat{N}}_{\tau}-N_{\tau})} | {\Phi}_{\tau}
\rangle } { \displaystyle  
\sum_{l_{\tau}=1}^L \langle {\Phi}_{\tau} |  
e^ {i {\varphi}_{l_{\tau}} ({\hat{N}}_{\tau}-N_{\tau})} | {\Phi}_{\tau}\rangle}, 
\label{yltau}
\end{equation}
and the  $\hat{O}$-operator overlap 
%
%%%%%%%%%%%%%%--------OVERLAP OF THE OPERATOR-------%%%%%%%%%%%%%%%%%%%%%%%
%
\begin{equation}
O({\varphi}_{\pi},{\varphi}_{\nu}) = 
\frac {\langle {\Phi} | \hat{O} 
e^ {i {\varphi}_{l_{\pi}}{\hat{N}}_{\pi}}
e^ {i {\varphi}_{l_{\nu}} {\hat{N}}_{\nu}} | {\Phi}
\rangle   }  
{\langle {\Phi} | 
e^ {i {\varphi}_{l_{\pi}} {\hat{N}}_{\pi}} 
e^ {i {\varphi}_{l_{\nu}} {\hat{N}}_{\nu}} 
 | {\Phi} \rangle }, 
\label{overO}
\end{equation}
with $| {\Phi} \rangle = | {\Phi}_{\pi} \rangle | {\Phi}_{\nu} \rangle $.
Another useful quantity is the norm overlap, which  is given by

%%%%%%%%%%%%%%%%--------OVERLAP OF THE NORM------%%%%%%%%%%%%%%%%%%%%%%

\begin{equation}
x_{l_\tau} = \langle {\Phi}_{\tau} | 
e^ {i {\varphi}_{l_{\tau}} ({\hat{N}}_{\tau}-N_{\tau})}
| {\Phi}_{\tau} \rangle,
\label{overN}
\end{equation}
clearly,  ${ y_{l_{\tau}} } $ and $x_{l_\tau}$ are related by
 ${\displaystyle y_{l_{\tau}} }  = {\displaystyle y^{\tau}_{0} }
x_{l_\tau}$,  with $y^{\tau}_{0} =( \sum_{l_{\tau}=1}^L x_{l_\tau})^{-1}$.

 As it will be illustrated for the Hamiltonian operator, any operator overlap 
can be calculated using the generalized Wick theorem \cite{Bal69}.
This theorem allows to express  matrix elements
of the form $  \langle \Phi | \hat{O}| \tilde {\Phi} \rangle$, 
exactly in the same way as the ordinary Wick theorem, as the sum of
all possible contracted products, where $  | {\Phi} \rangle$ and 
$  | \tilde {\Phi} \rangle$ are  quasiparticle vacua. Thus, in particular
\begin{eqnarray}
\langle \Phi |{c}^{\dagger}_{k_1}{c}^{\dagger}_{k_2}c_{k_4}c_{k_3} & & |\tilde {\Phi}\rangle = 
\langle \Phi  |\tilde {\Phi} \rangle^{-1} \left [
\langle \Phi |{c}^{\dagger}_{k_1}c_{k_3} |\tilde {\Phi} \rangle       
\langle \Phi |{c}^{\dagger}_{k_2}c_{k_4} |\tilde {\Phi} \rangle \right.\nonumber\\
 & &  -\left.\langle \Phi |{c}^{\dagger}_{k_1}c_{k_4} |\tilde {\Phi} \rangle
\langle \Phi |{c}^{\dagger}_{k_2}c_{k_3} |\tilde {\Phi} \rangle 
 +\langle \Phi |{c}^{\dagger}_{k_1}{c}^{\dagger}_{k_2}|\tilde {\Phi} \rangle
\langle \Phi |c_{k_4}c_{k_3} |\tilde {\Phi} \rangle \right ]
\label{extwick}
\end{eqnarray}
Looking at eq.~(\ref{PME}) we see that this theorem allows to express 
projected expectation values in terms of mean field matrix elements.
In eq.~(\ref{extwick}) the first term is the Hartree term, the second one the Fock term and
the last one the Bogoliubov term.
>From this expression we see that problems may appear when
$\langle \Phi | \tilde {\Phi} \rangle \approx 0$. This point will be discussed  
 later on and its consequences for the  effective force will be analyzed in 
 Appendix~\ref{appen:poles}. For the PNP method,
the basic building blocks of Wick's theorem 
 are the  $\varphi$-dependent generalized density  matrix and pairing
 tensors 
%
%%%%%%%%%%%%%DENSITY MATRIX AND PAIRING TENSOR%%%%%%%%%%%%%%%%%%%%%%%%%
%
%
\begin{eqnarray}
{\rho}_{kk'} (\varphi) &=& 
 \frac{ \langle \Phi | {c^{\dagger}_{k'}}c_k 
 | \tilde {\Phi} \rangle }{\langle \Phi | 
  \tilde {\Phi} \rangle } =  
 \frac{ \langle \Phi | {c^{\dagger}_{k'}}c_k 
e^{i\varphi\hat{N}} | \Phi \rangle }{\langle \Phi | 
e^{i\varphi\hat{N}} | \Phi \rangle } = 
{ \left ({\mathcal V}^*(\varphi)V^T \right)}_{kk'} 
\label{eq:rhophi} \\
{\kappa}_{kk'}^{10} (\varphi) &=& 
\frac{ \langle \Phi | c_{k'}c_k 
 | \tilde {\Phi} \rangle }{\langle \Phi | 
  \tilde {\Phi }\rangle } =
\frac{ \langle \Phi | c_{k'}c_k 
e^{i\varphi\hat{N}} | \Phi \rangle }{\langle \Phi | 
e^{i\varphi\hat{N}} | \Phi \rangle } =
 {  \left( {\mathcal V}^*(\varphi)U^T \right)}_{kk'}\label{eq:kap1phi} \\
{\kappa}_{kk'}^{01} (\varphi) &=& 
\frac{ \langle \Phi | c_{k}^{\dagger}
c_{k'}^{\dagger}  | \tilde{\Phi} \rangle }{\langle \Phi
 | \tilde{\Phi} \rangle } = 
\frac{ \langle \Phi | c_{k}^{\dagger}
c_{k'}^{\dagger} e^{ i\varphi\hat{N}} | \Phi \rangle }{\langle \Phi | \
e^{i\varphi\hat{N}} | \Phi \rangle } = 
{ \left( V {\mathcal U}^{\dagger}(\varphi) \right) }_{kk'}\label{eq:kap0phi}
\end{eqnarray}
 where we have used the notation
$|\tilde{\Phi}> = e^{i\varphi\hat{N}} |\Phi>$, the indices $(k, k')$
belonging to the same isospin channel $\tau$. 
The matrices ${\mathcal U} (\varphi)$ and  ${\mathcal V} (\varphi)$ are related
to $U$ and $V$ of eq.~(\ref{BogT}) by
%
%%%%%%%%%%%%%%%NEW MATRIXES U(PHI) AND V(PHI)%%%%%%%%%%%%%%%%%%%%%%%%%%
%
\begin{equation}
{\mathcal U} (\varphi) \, = \, U\,+\,V^*{{\mathcal X} (\varphi)}^* \;\;\;\;\;\;\;\;
{\mathcal V} (\varphi) \, = \, V\,+\,U^*{{\mathcal X} (\varphi)}^*,
\end{equation}
and the contraction ${\mathcal X} (\varphi)$  is given by
%
%%%%%%%%%%%%%%%-------X(PHI) MATRIX------------%%%%%%%%%%%%%%%%%%%%%%%%%%%
%
\begin{equation}
{\mathcal X}_{\mu \nu} (\varphi) 
= \frac{ \langle \Phi | {\alpha}_{\nu} {\alpha}_{\mu} 
 | \tilde{\Phi} \rangle }{\langle \Phi  
 | \tilde{\Phi} \rangle } = { \left (
T_{21}^* (\varphi) T_{22}^{-1} (\varphi)\right) }_{\mu \nu},
\label{Xmunu}
\end{equation}
here again the indices ${\mu, \nu}$ belong to the same isospin channel and
%%%%%%%%%%%%%%--------T21 AND T22--------------%%%%%%%%%%%%%%%%%%%%%%%%%%%
%
\begin{eqnarray}
T_{22} \, ({\varphi}) & = & \cos({\varphi}) \cdot
{\ide} \, - \, i \sin ({\varphi}) \cdot {\hat{N}}^{11}  \\
T_{21}  (\varphi) \,& = & \, - \, i \sin ({\varphi}) \cdot {\hat{N}}^{20}
\end{eqnarray}
with ${N}^{11}=U^{\dagger}U-V^{\dagger}V$ and
${N}^{20}=U^{\dagger}V^*-V^{\dagger}U^*$. The matrices ${N}^{11}$ and ${N}^{20}$ 
represent the number operator in the quasiparticle basis 
of eq.(\ref{BogT}).

The nuclear Hamiltonian can be written as 
\begin{eqnarray}
\hat{H} & = & T + \hat{V}_{DI} + \hat{V}_{DD} \nonumber \\
        & = & \sum_{k_1k_2} t_{k_1k_2}{c}^{\dagger}_{k_1}c_{k_2} \, + \,
\frac{1}{4}
\sum_{k_1k_2k_3k_4}{\bar{v}}_{k_1k_2k_3k_4}{c}^{\dagger}_{k_1}
{c}^{\dagger}_{k_2}c_{k_4}c_{k_3} +\hat{V}_{DD}
\end{eqnarray}
where the density independent part, $\hat{V}_{DI}$, represents the nuclear 
interaction (Coulomb included) with
the exception of a possible density-dependent term (the well known density
dependent term of the Skyrme and Gogny forces) which  is represented by
$ \hat{V}_{DD}$. In our case $ \hat{V}_{DD}$ is given by the last term of 
eq.~(\ref{eq:vgog}).

\subsection{ Particle Number Projected Energy for the non-density
dependent parts of the interaction.}
With the definitions above   we can formulate the particle number
projected theory for  non--separable interactions, like the Gogny 
force~(Appendix~\ref{appen:GogF}), in a similar way as in the mean field 
approximation. Using
eq.~(\ref{PME}) and the generalized Wick theorem, the density independent part 
of the projected energy is given by
\begin{eqnarray}
 E^P_{DI} & = &  \sum_{l_{\pi}=1}^L \sum_{l_{\nu}=1}^L
y_{l_{\pi}} y_{l_{\nu}}
\cdot   Tr  \biggl \{t  \left ( 
{\rho}({\varphi}_{l_{\pi}}) + 
{\rho}({\varphi}_{l_{\nu}}) \right )   \nonumber \\ 
& + &  \frac{\,1\,}{2} \Bigl( {\Gamma}^{\pi \pi}
({\varphi}_{l_{\pi}}) \rho ({\varphi}_{l_{\pi}}) + {\Gamma}^{\pi \nu}
({\varphi}_{l_{\nu}}) \rho ({\varphi}_{l_{\pi}}) + {\Gamma}^{\nu \pi}
({\varphi}_{l_{\pi}}) \rho ({\varphi}_{l_{\nu}}) \nonumber  \\
& + &    {\Gamma}^{\nu \nu}
({\varphi}_{l_{\nu}}) \rho ({\varphi}_{l_{\nu}})  
-  
{\Delta}^{10,\pi}
({\varphi}_{l_{\pi}}){\kappa}^{01}({\varphi}_{l_{\pi}}) - {\Delta}^{10,\nu}
({\varphi}_{l_{\nu}}){\kappa}^{01}({\varphi}_{l_{\nu}}) \Bigr )
\biggr \} \, ,
\end{eqnarray}
where the traces run over the configuration space, here assumed to be the
same for protons and neutrons, and 
\begin{eqnarray}
{\Gamma}^{\tau {\tau}'}_{k_1k_3} ({\varphi}_{l_{{\tau}'}}) &= & \sum_{k_2k_4} 
{\bar{v}}_{k_1k_2k_3k_4} {\rho}_{k_4k_2} ({\varphi}_{l_{{\tau}'}}), \\ 
{\Delta}^{10,\tau}_{k_1k_2} (\varphi_{\tau}) &=& \frac{1}{2} \sum_{k_3k_4} 
{\bar{v}}_{k_1k_2k_3k_4}{\kappa}^{10}_{k_3k_4} (\varphi_{\tau}).
\end{eqnarray}
In ${\Gamma}^{\tau {\tau}'}$ the indices $({k_1,k_3})$ belong to the isospin
$\tau$ and $({k_2,k_4})$ to $\tau'$, in  ${\Delta}^{10,\tau}$ all indices belong 
to the isospin $\tau$~\footnote{Notice that the pairing field
 $\Delta$ does not mix isospin.}. 
Since $\sum_{l_{\tau}=1}^L y_{l_{\tau}}=1$, we can write the proton or neutron
projected density as
\begin{eqnarray}
 {\rho}^{P\tau}_{k,k'} =
  \frac{\langle \Phi| {c_{k'}}^{\dagger}c_k  \hat{P}^{N_{\tau}}  | \Phi \rangle } 
  { \langle \Phi |\hat{P}^{N_{\tau}} | \Phi \rangle} =
 \sum_{l_{\tau}=1}^L y_{l_{\tau}}
{\rho}^{\tau}_{k,k'}({\varphi}_{l_{\tau}})  \, ,
\label{eq:rhop}
\end{eqnarray}  
with $(k,k')$ belonging to the isospin $\tau$, and the projected energy as
\begin{eqnarray}
E^P_{DI} & =  & Tr \Biggl \{ 
\Bigl ( t^{\pi} + \frac{\,1\,}{2} {\Gamma}^{\pi,P\nu}\Bigr ){\rho}^{P\pi} +
   \Bigl ( t^{\nu} + \frac{\,1\,}{2}{\Gamma}^{\nu,P\pi}\Bigr ){\rho}^{P\nu}
 + \nonumber \\
& + & \frac{\,1\,}{2} \, \sum_{\tau} \sum_{l_{\tau}=1}^L y_{l_{\tau}} 
\Bigl ({\Gamma}^{\tau \tau}
({\varphi}_{l_{\tau}}) \rho ({\varphi}_{l_{\tau}}) -
{\Delta}^{10,\tau}({\varphi}_{l_{\tau}}){\kappa}^{01}({\varphi}_{l_{\tau}})
 \Bigr )\Biggr \}  \, ,
 \label{eq:epdi}
\end{eqnarray} 
where 

\begin{equation}
{\Gamma}^{\tau, P{\tau}'}_{k_1k_3}=\sum_{l_{{\tau}'}=1}^L y_{l_{{\tau}'}}
{\Gamma}^{\tau
{\tau}'}_{k_1k_3} ({\varphi}_{l_{{\tau}'}})=
\sum_{k_2k_4} 
{\bar{v}}_{k_1k_2k_3k_4} {\rho}^{P{\tau}'}_{k_4k_2} \, .
\end{equation} 
and $ \tau \ne \tau'$.
By $(P{\tau}')$ in ${\Gamma}^{\tau, P{\tau}'}$ and in some notation to be
introduced later on (~see also eq.~\ref{eq:rhop}~) we want to indicate that 
these quantities are ${\varphi}_{l_{{\tau}'}}$
independent, i.e., the integration  on ${\varphi}_{l_{{\tau}'}}$
has been performed.
In these fields, the indices ${(k_1,k_3)}$ belong
to the isospin $\tau$. The expressions (\ref{eq:rhop}) and (\ref{eq:epdi})
can be rewritten taking into account that
\begin{equation}
y_{L-l} = y_l^*,  \quad
\rho (\varphi_{L-l} ) =\rho^{\dagger} (\varphi_{l} ), \quad
\kappa^{10} (\varphi_{L-l} ) =  e ^{-2 i\varphi_{l}} \kappa^{01} 
(\varphi_{l} )^*.
\label{eq:DESCsum}
\end{equation}
In this way we can simplify the sum in eq.~(\ref{eq:rhop}) for the projected density
to\footnote{Notice that the sum in eq.~(\ref{Fom-exp}) from $l=1$ to $L$ can be
written as a sum from $l=0$ to $L-1$.}

\begin{eqnarray}
{\rho}^{P\tau} & = & 
 Re \left \{   y_0^{\tau} \sum_{l_{\tau}=0}^{[L/2]} {\mathcal C}_{l_{\tau}} x_{l_{\tau}} 
{\rho}^{\tau}({\varphi}_{l_{\tau}})  \right \} \, ,
\end{eqnarray}
where we have taken into account that 
$y_{l_{\tau}}=y_0^{\tau}\cdot x_{l_{\tau}}$. By $[L/2]$ we represent the 
integer part of $L/2$ and
\begin{eqnarray}
{\mathcal C}_{l_{\tau}} = \left\{ \begin{array}{lll}
               1 & \mbox{if $l_{\tau}=0$      } \\
	       1 & \mbox{if $l_{\tau}= L/2$ and L even  } \\
	       2 & \mbox{otherwise.                   }
	       \end{array}
	       \right .
\label{eq:C0}
\end{eqnarray}
 In the same way, for the projected energy,  eq.~(\ref{eq:epdi}), we obtain
\begin{eqnarray}
E^P_{DI} & =  & Tr \Biggl \{ 
\Bigl ( t^{\pi} + \frac{\,1\,}{2} {\Gamma}^{\pi,P\nu}\Bigr ){\rho}^{P\pi} +
   \Bigl ( t^{\nu} + \frac{\,1\,}{2}{\Gamma}^{\nu,P\pi}\Bigr ){\rho}^{P\nu}
  \nonumber \\
 & + & 
    Re \sum_{\tau} y_0^{\tau} \,
  \sum_{l_\tau=0}^{[L/2]} 
x_{l_{\tau}} \frac{\,{\mathcal C}_{l_{\tau}}\,}{2} \Bigl ( {\Gamma}^{\tau \tau}
({\varphi}_{l_{\tau}}) \rho ({\varphi}_{l_{\tau}}) -
{\Delta}^{10,\tau}({\varphi}_{l_{\tau}}){\kappa}^{01}({\varphi}_{l_{\tau}})
 \Bigr )  \Biggr \}. 
\label{eq:EP2}
\end{eqnarray}

\section{Particle Number Projection for the density dependent part of the 
interaction}
\label{sec:DDP}
Density dependent interactions have a term which simulates a G-matrix 
through an explicit dependence of the nuclear density. This term, $V_{DD}$,
 was introduced in the mean field  approximation. In the MFA only
{\it diagonal} matrix elements between product wave functions are needed 
to calculate the energy and, consequently, $V_{DD}$ is constructed to depend
on the mean field density.
  In theories beyond mean field, the density dependent part is given by
\begin{eqnarray}
E^P_{DD}   & = & \frac{\langle \Psi_N | {\hat{V}}_{DD}\left [ \overline{\rho} 
(\vec{r}) \right ] | \Psi_N \rangle} 
{\langle \Psi_N | \Psi_N \rangle} \nonumber \\
& = & \frac{ {\displaystyle \int} d {\varphi}_{\pi} {\displaystyle 
\int }   d {\varphi}_{\nu} \langle 
\Phi  | {\hat{V}}_{DD}
 \left [ \overline{\rho} (\vec{r}) \right ] 
e^{i{\varphi}_{\pi} {\hat{N}}_{\pi}} 
e^{i{\varphi}_{\nu} {\hat{N}}_{\nu} } | \Phi \rangle  } 
{{\displaystyle \int } d {\varphi}_{\pi} {\displaystyle  \int } 
 d {\varphi}_{\nu} \langle 
\Phi | e^{i{\varphi}_{\pi} {\hat{N}}_{\pi}} 
e^{i{\varphi}_{\nu} {\hat{N}}_{\nu} } | \Phi \rangle }
\label{eq:epdd}
\end{eqnarray}
where $\left [ \overline{\rho} (\vec{r}) \right ]$ indicates the explicit
dependence of $V_{DD}$ on a density $ \overline{\rho} (\vec{r})$ to be specified. Looking
at these expressions it is not obvious which dependence should be used.
There are two more or less straightforward prescriptions \cite{VER.00}
for $ \overline{\rho} (\vec{r})$~:
\begin{enumerate}

\item The first prescription is inspired by the following consideration~:
In the MFA, the energy is given by
$\langle \Phi | \hat{H} | \Phi \rangle /{\langle \Phi | \Phi \rangle}$
and $V_{DD}$ is assumed to  depend on the density  
$\langle \Phi | \hat{\rho} | \Phi \rangle/{\langle \Phi | \Phi \rangle}$.
On the other hand, if the wave function which describes the nuclear system 
is the projected wave function $| \Psi_N \rangle$,   we 
have to calculate the matrix element
$\langle \Psi_N | \hat{V}_{DD} | \Psi_N \rangle /{\langle \Psi_N | \Psi_N \rangle}$
(see the middle term in eq.~(\ref{eq:epdd})).
It seems reasonable, therefore,  to use in $V_{DD}$ the density 
$ \overline{\rho} (\vec{r})= \langle \Psi_N | \hat{\rho} | \Psi_N \rangle /
{\langle \Psi_N | \Psi_N \rangle} = {\rho}^{P\tau}(\vec{r})$, i.e. the projected 
density. 

\item The other prescription has been guided by the choice usually done in the
Generator Coordinate method with density dependent forces~\cite{Bon90}. The
philosophy behind this prescription is the following: to
evaluate eq.~(\ref{eq:epdd}) we have to calculate matrix elements between different
product wave functions $|\Phi \rangle$ and $| \tilde{\Phi} \rangle$
($|\tilde{\Phi} \rangle =e^{i\varphi\hat{N}}| \Phi \rangle$)
(see last term in eq.~(\ref{eq:epdd})). 
Then, to calculate  matrix elements of the form
\begin{equation}
\frac{\langle \Phi | {\hat{V}}_{DD} | \tilde{\Phi} \rangle}{\langle \Phi |
\tilde{\Phi} \rangle}
\end{equation}
we choose the mixed density
\begin{equation}
  \overline{\rho} (\vec{r})= {\rho}_{\varphi} (\vec{r}) =
 \frac{\langle \Phi | \hat{\rho} (\vec{r})| \tilde{\Phi} \rangle}{\langle \Phi |
\tilde{\Phi} \rangle}
\end{equation}
to be used in $ {\hat{V}}_{DD}$.
We shall call this approach the mixed density prescription.
\end{enumerate}
Both prescriptions have been tested with the Gogny force in the Lipkin
Nogami approach \cite{VER.97} and practically no difference was found in the 
numerical applications. One should notice that in the second prescription 
$\overline{\rho} (\vec{r})$ depends on the angle $\varphi$ at variance with
prescription 1.

We shall now proceed to evaluate the projected energy with both prescriptions.

\subsection{Prescription 1: Projected Density.}

Using the projected density prescription, the projected energy of the 
density--dependent term
($E^P_{DD}$) is obtained in a simple way. One just has to use in $ {\hat{V}}_{DD}$
the projected density. In the  case of the Gogny force
 (see Appendix~\ref{appen:GogF}) which we are 
considering here, the value  $x_0 =1$ for the spin exchange  parameter 
guarantees that 
there is no contribution from $ {\hat{V}}_{DD}$ neither to 
the pairing field $\Delta$ nor to the Hartree-Fock field of the type 
$\Gamma^{\tau\tau}$, i.e., without isospin mixing.
We find
\begin{eqnarray}
E^P_{DD} & = & \frac{1}{2} Tr  \left
 ( {\breve{\Gamma}}^{\nu,P\pi} \cdot {\rho}^{P\nu }+{\breve{\Gamma}}^{\pi,P\nu} 
 \cdot {\rho}^{P\pi }\right )\; ,
\label{eq:EPDD1}
\end{eqnarray}
 where
\begin{eqnarray}
{\breve{\Gamma}}^{\tau,P\tau'}_{k_1 k_3} & = & \sum_{k_2k_4} { \left ( \displaystyle
 { {\bar{v}}_{DD} }
[ {\rho}^P ] \right ) }_{k_1k_2k_3k_4} {\rho}_{k_4 k_2}^{P\tau'}\; ,
\end{eqnarray}
with $\tau \ne \tau'$ and  
\begin{eqnarray}
{\left ( { {\bar{v}}_{DD} } [ {\rho}^P ] \right ) }_{k_1k_2k_3k_4} = 
\langle k_1k_2 | V_{DD} [ {\rho}^P ] |k_3k_4 \rangle -
\langle k_1k_2 | V_{DD} [ {\rho}^P ] |k_4k_3 \rangle \; ,
\end{eqnarray}

the symbol $\,\breve{}\,$ is used to specify that we are dealing with the DD part
of the interaction.
\subsection{Prescription 2: Mixed Density.}

We assume in this section that $V_{DD}$ depends on the mixed density 
\begin{eqnarray}
{\rho}_{{\varphi}_{\pi}{\varphi}_{\nu}} (\vec{r}) &  =  & 
\frac { \langle \Phi | \hat{\rho} 
(\vec{r}) e^{i{\varphi}_{\pi} {\hat{N}}_{\pi}} | \Phi \rangle} 
{ \langle \Phi | e^{i{\varphi}_{\pi} {\hat{N}}_{\pi}} | \Phi
\rangle} + 
\frac { \langle \Phi | \hat{\rho} 
(\vec{r}) e^{i{\varphi}_{\nu} {\hat{N}}_{\nu}} | \Phi \rangle} 
{ \langle \Phi | e^{i{\varphi}_{\nu} {\hat{N}}_{\nu}} | \Phi 
\rangle} 
\label{mixdens}
\end{eqnarray}

The contribution of  $\hat{V}_{DD} $ to the energy  is given by 
eq.~(\ref{eq:epdd}) with $\left [ \overline{\rho} (\vec{r}) \right ]=
{\rho}_{{\varphi}_{\pi}{\varphi}_{\nu}} (\vec{r}) $.
Since the mixed density depends on the angles, we cannot simplify the double
integral to a single one as we did before. 
Replacing the integrals by sums and using~(\ref{eq:DESCsum}) again, we obtain
\begin{eqnarray}
E^{P}_{DD} & = & \frac{1} {2} \sum_{\tau \ne \tau'}\left \{   {\displaystyle Re } \left (  
\sum_{l_{\tau}=0}^{[L/2]} y_{l_{\tau}} \sum_{l_{\tau'}=0}^{[L/2]} y_{l_{\tau'}}
 {\mathcal C}_{l_\tau l_{\tau'}} \cdot {\displaystyle Tr } \Bigl (
{\breve{\Gamma}}^{\tau \tau'} \bigl [ \rho({\varphi}_{l_{\tau}}),
\rho({\varphi}_{l_{\tau'}}) \bigr ] \cdot \rho ({\varphi}_{l_{\tau}}) \Bigr )
\right )    \nonumber \right. \\
 &  + & \hspace{-0.1cm} \left. 2\cdot {\displaystyle Re} \left 
( \sum_{l_{\tau}=1}^{[L/2]-1} y_{l_{\tau}} 
\sum_{l_{\tau'}=1}^{[L/2]-1} y^*_{l_{\tau'}} {\displaystyle Tr } \Bigl (
{\breve{\Gamma}}^{\tau \tau'} \bigl [ \rho({\varphi}_{l_{\tau}}),
\rho^*({\varphi}_{l_{\tau'}}) \bigr ] \cdot \rho ({\varphi}_{l_{\tau}}) \Bigr )
\right ) 
\right \}
\label{eq:EPDD2}
\end{eqnarray}
where  $[L/2]$ represents, as before,  the integer part of   $L/2$ and
\begin{equation}
\mathcal C_{l_\tau l_{\tau'}} = \left\{ \begin{array}{lll}
               1 & \mbox{if $l_{\tau}=0$ or $L/2$ and $l_{\tau'}=0$ or $L/2$ ($L$ even) } \\
	       1 & \mbox{if $l_{\tau}=0$ and $l_{\tau'}=0$  ($L$ odd) } \\
	       2 & \mbox{otherwise.                   }
	       \end{array}
	       \right .
\label{eq:C}
\end{equation}	       

In this expression we have furthermore introduced the following notation

\begin{eqnarray}
{\breve{\Gamma}}^{\tau \tau'}_{k_1 k_3} 
 \bigl [ \rho({\varphi}_{l_{\tau}}),
\rho({\varphi}_{l_{\tau'}}) \bigr ] & = & \sum_{k_4 k_2} 
{ \left ( \bar{v}_{DD}  \left 
[ \rho({\varphi}_{l_{\tau}}),\rho({\varphi}_{l_{\tau'}})
\right ] \right ) }_{k_1 k_2 k_3 k_4 } 
\rho_{k_4 k_2} ({\varphi}_{l_{\tau'}}) \nonumber \\ 
{\breve{\Gamma}}^{\tau \tau'}_{k_1 k_3} 
 \bigl [ \rho({\varphi}_{l_{\tau}}),
\rho^*({\varphi}_{l_{\tau'}}) \bigr ] & = & \sum_{k_4 k_2} 
{ \left ( \bar{v}_{DD}  \left 
[ \rho({\varphi}_{l_{\tau}}),\rho^*({\varphi}_{l_{\tau'}})
\right ] \right ) }_{k_1 k_2 k_3 k_4 } 
\rho^*_{k_4 k_2} ({\varphi}_{l_{\tau'}}). 
\label{eq:gammatil} 
\end{eqnarray}
Here again the indices $({k_1,k_3})$ belong to the isospin
$\tau$ and $({k_2,k_4})$ to $\tau'$ $(\tau~\ne~\tau')$
and ${ \left ( \bar{v}_{DD}  \left 
[ \rho({\varphi}_{l_{\tau}}),\rho({\varphi}_{l_{\tau'}})
\right ] \right ) }_{k_1 k_2 k_3 k_4 }$ 
 is the density--dependent matrix element calculated with the
density $\rho^{\tau}(\varphi_{\tau},\vec{r}) + 
\rho^{\tau'}(\varphi_{\tau'},\vec{r})$  in $V_{DD}$,
similarly, ${ \left ( \bar{v}_{DD}  \left 
[ \rho({\varphi}_{l_{\tau}}),\rho^*({\varphi}_{l_{\tau'}})
\right ] \right ) }$  is calculated with 
$\rho^{\tau}(\varphi_{\tau},\vec{r}) + 
(\rho^{\tau'}(\varphi_{\tau'},\vec{r}))^*$ in $V_{DD}$.

If we compare the expression~(\ref{eq:EPDD2}) with the result obtained for 
$E^{P}_{DD}$ with the projected prescription~(\ref{eq:EPDD1}), it is easy to
observe that~(\ref{eq:EPDD2}) implies a more elaborated calculation. With the
projected prescription, we only have to calculate the projected density $\rho^P$,
the field ${\breve{\Gamma}} [ \rho^P ]$ and then one  easily gets
 $E^P_{DD}$. But in
order to evaluate  expression~(\ref{eq:EPDD2}) it is necessary to calculate
 $  ( L^2/2 + 2 ) $ different fields. For example, for $L=8$, we have to calculate 
 $34$ fields. 
\section{Restoration of the particle number symmetry before the variation 
in HFB theories with non-separable forces.}
\label{sec:VAP}

Using  equation~(\ref{eq:EP2}), we can perform a 
restoration of the particle number symmetry, projecting after the variational 
Hartree--Fock--Bogoliubov equations have been solved. In this case the HFB
wave function is not self--consistently determined.  In particular, if the
HFB calculation collapses to the pairing uncorrelated HF one, the PNP method
does not provide a better approximation.
In this section we shall derive the projected variational equations in order to
solve the self--consistent problem, that means, the variation after the
projection method. According to the Ritz variational principle, the intrinsic
wave function $| \Phi \rangle $ has to be determined by minimization of the 
projected energy, i.e.,
\begin{equation}
\delta E^P = \delta \; \frac{\langle \Phi | \hat{H} \hat{P}^{N_{\pi}} \hat{P}^{N_{\nu}}
|\Phi \rangle  }{\langle \Phi |\hat{P}^{N_{\pi}} 
\hat{P}^{N_{\nu}}|\Phi \rangle} =0
\end{equation}

The variational Hilbert space generated by product wave functions can be 
parametrized by the Thouless theorem~\cite{RS.80}, which allows to write 
any product wave function, $|\Phi \rangle$, in the form
\begin{equation}
 |\Phi\rangle= \mathcal{N} \exp{ (\frac{1}{2} \sum_{\mu  \nu} C_{\mu
\nu}{\alpha}_{\mu}^{\dagger}{\alpha}_{\nu}^{\dagger} )}|{\Phi}_0\rangle 
\end{equation}
where $|\Phi_0 \rangle$ is a reference product wave function non--orthogonal to
$|\Phi \rangle$ and $C_{\mu \nu}$ are the variational parameters.
We shall first study the variation of the density independent part of the 
interaction, eq.~(\ref{eq:EP2}).
An arbitrary variation of the energy is given by
\begin{eqnarray}
{\delta E}^P_{DI} & = &  \frac{1}{2} \sum_{\mu  \nu} 
{ \partial E^P_{DI} \over \partial C^*_{\mu \nu}}\ \delta C^*_{\mu \nu} 
+ \, {  \partial E^P_{DI} \over \partial C_{\mu \nu} }\
 \delta C_{\mu \nu}  \nonumber \\ 
& = & \frac{1}{2} \sum_{\mu \nu} \left( \frac{\langle   \Phi | 
{\alpha}_{\nu} {\alpha}_{\mu}( \hat{H}_{DI}-E^P_{DI}) 
{\hat{P}}^N  |  \Phi  \rangle }
{\langle\, \Phi  |  {\hat{P}}^N  |  \Phi \, \rangle }\ 
\delta C^*_{\mu \nu} + h.c. \right) \nonumber \\
& =& \frac{1}{2}\sum_{\mu \nu} \left( {\mathcal H}^{20,P}_{\mu \nu}\ 
\delta C^*_{\mu \nu} + h.c. \right)
\end{eqnarray}

where we have introduced the projected gradient 
${\mathcal H}^{20,P}_{\mu \nu}$. Here the indices $(\mu, \nu)$ run over 
the proton and neutron configuration space.
The solution of the projected variational problem  can be obtained by 
the conjugate gradient method~\cite{Egi95}.

Using the generalized Wick theorem~\cite{Bal69}, ${\mathcal H}^{20,P}_{\mu \nu}$
can be written as
\begin{eqnarray}
{\mathcal H}^{20,P}_{\mu \nu,\tau}  = 
  \sum_{l_{\tau}=1}^{L} 
y_{l_{\tau}} \Biggl \{ {\mathcal H}^{20}_{\mu \nu,\tau} ({\varphi}_{l_{\tau}})  
+  {\mathcal E}^{\tau}({\varphi}_{l_{\tau}}) 
\left ( {\mathcal X}_{\mu \nu}({\varphi}_{l_{\tau}}) - {\mathcal X}^{P\tau}_{\mu \nu} \right ) 
\Biggr \} 
\end{eqnarray}
with the index $\tau$ corresponding to the isospin of the quasiparticle 
operators  $ (\alpha_\mu, \alpha_\nu)$. We have furthermore introduced
${\mathcal X}^{P\tau}  =  \sum_{l_{\tau}} y_{l_{\tau}} {\mathcal X}
({\varphi}_{l_{\tau}})$ and
\begin{eqnarray}
{\mathcal H}^{20}_{\mu \nu,\tau} ({\varphi}_{l_{\tau}})& &    = 
  \left ( {\mathcal U}^{\dagger}({\varphi}_{l_{\tau}}) h({\varphi}_{l_{\tau}})
{\mathcal V}^* ({\varphi}_{l_{\tau}})
 -  {\mathcal V}^{\dagger}({\varphi}_{l_{\tau}}) h^T({\varphi}_{l_{\tau}}) 
 {\mathcal U}^* ({\varphi}_{l_{\tau}})  \right.   \nonumber \\
& & \left.   + \,  {\mathcal U}^{\dagger}({\varphi}_{l_{\tau}}) 
{\Delta}^{10,\tau}({\varphi}_{l_{\tau}})
{\mathcal U}^* ({\varphi}_{l_{\tau}})
-   {\mathcal V}^{\dagger}({\varphi}_{l_{\tau}}) 
 {\Delta}^{01,\tau}({\varphi}_{l_{\tau}}) 
 {\mathcal V}^* ({\varphi}_{l_{\tau}})  \right )_{\mu \nu }
\end{eqnarray}
with
\begin{eqnarray}
h({\varphi}_{l_{\tau}}) & = & t + {\Gamma}^{\tau,P{\tau}'} +  
{\Gamma}^{\tau \tau}({\varphi}_{l_{\tau}}),
\end{eqnarray}
again $\tau \ne \tau'$, and  
\begin{eqnarray}
 {\mathcal E}^{\tau}&&({\varphi}_{l_{\tau}})  = E^{\tau}_{DI}({\varphi}_{l_{\tau}}) 
  \nonumber \\
 &&+ \frac{1}{2} Tr \Biggl \{ \left (  V^T{\left (t + {\Gamma}^{\tau,P\tau'} \right )}^T U - 
U^T \left ( t+{\Gamma}^{\tau,P\tau'} \right )V  \right ) 
\left ( {\mathcal X}({\varphi}_{l_{\tau}}) - {\mathcal X}^{P\tau} \right ) \Biggr \}  
\end{eqnarray}
where
\begin{eqnarray} 
E^{\tau}_{DI}({\varphi}_{l_{\tau}})  =  \frac{1}{2}Tr \Bigl ( {\Gamma}^{\tau \tau} 
({\varphi}_{l_{\tau}}) {\rho}^{\tau}({\varphi}_{l_{\tau}})\,-\,
{\Delta}^{10,\tau} ({\varphi}_{l_{\tau}}) {\kappa}^{01,\tau}({\varphi}_{l_{\tau}}) \Bigr )
\end{eqnarray}

Taking into account eq.~(\ref{eq:DESCsum}), we
can simplify the gradient expression to the region $[0,L/2]$ in the following
way~:
\begin{eqnarray}
{\mathcal H}^{20,P}_{\mu \nu,\tau} & = &  
   y^{\tau}_0  Re \sum_{l_{\tau}=0}^{[L/2]} C_{l_{\tau}} x_{l_{\tau}} \left\{ 
{\mathcal H}^{20}_{\mu \nu,\tau} ({\varphi}_{l_{\tau}})  
+  {\mathcal E}^{\tau}({\varphi}_{l_{\tau}}) 
\left ( {\mathcal X}_{\mu \nu}({\varphi}_{l_{\tau}}) - {\mathcal X}^{P\tau}_{\mu \nu} \right )   
 \right\}. 
\label{eq:H20PsinDD}
\end{eqnarray}
 These expressions apply to the non-density dependent part of the interaction.
 To calculate the gradient of the density dependent part of the interaction we
 proceed as with the evaluation of the energy, accordingly we shall
 distinguish the two prescriptions adopted.  

\subsection{Prescription 1: Projected Density.}

In a similar way as in the case of  the density independent part of the 
interaction, the variation
of ${E}^P_{DD}$ is defined by ${\delta E}^P_{DD} = \frac{1}{2}
\sum_{\mu \nu} \left( \breve{{\mathcal H}}^{20,P}_{\mu \nu}\ \delta C^*_{\mu \nu} + h.c.\right)$.
Here the expression of $\breve{{\mathcal H}}^{20,P}_{\mu \nu}$ is somewhat 
more complicated due to the explicit dependence on the density of the 
interaction~:
\begin{eqnarray}
\breve{{\mathcal H}}^{20,P}_{\mu \nu, \tau}  =  \frac{\langle \,  \Phi \,|\, 
{\alpha}_{\nu} {\alpha}_{\mu}( {\hat{V}}_{DD}-E^P_{DD} ) 
{\hat{P}}^N \, |\,  \Phi \, \rangle }
{\langle\, \Phi \, | \, {\hat{P}}^N \, | \, \Phi \, \rangle }  + 
\frac{\langle \,  \Phi \,|\, 
\displaystyle { \frac{ \partial {\hat{V}}_{DD} }
{\partial {\rho}^{P\tau}} \frac{ \partial {\rho}^{P\tau}} 
{\partial C^*_{\mu \nu}} } 
{\hat{P}}^N \, |\,  \Phi \, \rangle }
{\langle\, \Phi \, | \, {\hat{P}}^N \, | \, \Phi \, \rangle }.
\label{h20pdd}
\end{eqnarray}
Using eq.~(\ref{rhodef}) one can calculate $\displaystyle {\frac{ \partial {\rho}^{P\tau} }
{ \partial C^*_{\mu \nu }} }$  very easily~:
\begin{equation}
\frac {\partial {\rho}^{P\tau} (\vec{r})}{\partial C^*_{\mu \nu }}  = 
\frac{\langle \Phi |{\alpha}_{\nu} {\alpha}_{\mu}
\left ( \sum_{k,k'} f_{kk'}(\vec{r}) c_k^{\dagger} c_{k'}  -{\rho}^{P\tau}
  (\vec{r}) \right ) {\hat{P}}^N | \Phi \rangle } 
{ \langle\, \Phi \, | \, {\hat{P}}^N \, | \, \Phi \, \rangle }.
\label{rhoptau}  
\end{equation}
  Using the notation $\delta 
{\breve{\Gamma}}^R_{kk'}$  for the rearrangement term 
\begin{equation}
{\delta \breve{\Gamma} }_{kk'}^R =  \frac{\langle \,  \Phi \,|\, 
\displaystyle { \frac{ \partial {\hat{V}}_{DD} }
{\partial {\rho}^{P\tau}} } f_{kk'} (\vec{r}) 
{\hat{P}}^N \, |\,  \Phi \, \rangle }
{\langle\, \Phi \, | \, {\hat{P}}^N \, | \, \Phi \, \rangle }
\label{rearr}
\end{equation}
where the indices $(k,k')$ and $(\mu,\nu)$ in eqs.~(\ref{rhoptau},\ref{rearr})
belong to the isospin $\tau$,
we can write the projected gradient of $E_{DD}$ in the following way~:
\begin{eqnarray}
\breve{{\mathcal H}}^{20,P}_{\mu \nu,\tau}  =  
y^{\tau}_0 Re\sum_{l_{\tau}=0}^{[L/2]} C_{l_{\tau}}
x_{l_{\tau}}  \Biggl \{ \breve{{\mathcal H}}^{20}_{\mu \nu,\tau} 
({\varphi}_{l_{\tau}}) 
 +    {\breve{\mathcal E}}^{\tau}({\varphi}_{l_{\tau}})  
   \Bigl ({\mathcal X}_{\mu \nu}({\varphi}_{l_{\tau}}) - 
{\mathcal X}^{P\tau}_{\mu \nu} \Bigr )  \Biggr \}
\label{eq:H20PconDD}
\end{eqnarray}
where
\begin{eqnarray}
\breve{\mathcal H}^{20}_{\mu \nu,\tau} ({\varphi}_{l_{\tau}}) & = &  
\left ( {\mathcal U}^{\dagger}({\varphi}_{l_{\tau}}) 
\left ( \breve{\Gamma}^{\tau,P\tau'}  +  \delta {\breve{\Gamma}}^R  \right ) 
{\mathcal V}^* ({\varphi}_{l_{\tau}}) \right. \nonumber \\
     & - &    \left. {\mathcal V}^{\dagger} ({\varphi}_{l_{\tau}})
{ \left ( \breve{{\Gamma}}^{\tau,P\tau'} + \delta {\breve{\Gamma}}^R  \right ) }^T  
{\mathcal U}^* ({\varphi}_{l_{\tau}})  \right )_{\mu \nu} 
\end{eqnarray}
and 

\begin{eqnarray}
 {\breve{\mathcal E}}^{\tau}({\varphi}_{l_{\tau}})  & = & \frac{1}{2}
Tr \Biggl \{ \left (  V^T{\left ( \breve{\Gamma}^{\tau,P\tau'} + \delta {\breve{\Gamma}}^R  
\right )}^T U \right. \nonumber \\ 
& - &\left. 
U^T \left ( \breve{\Gamma}^{\tau,P\tau'}  + \delta {\breve{\Gamma}}^R  \right )V  \right )
 \left ( {\mathcal X}({\varphi}_{l_{\tau}}) - {\mathcal X}^{P\tau } \right ) 
\Biggr \}  
\end{eqnarray}
This is a similar expression to the projected gradient of the  non
density--dependent part,~(\ref{eq:H20PsinDD}), they differ only by the term
$E^{\tau} ({\varphi}_{l_{\tau}})$, which  does not contribute in this case 
because the specific value of the coefficient $x_0$ in the Gogny force 
makes it zero.

\subsubsection{Prescription 2: Mixed Density.}

 We shall now derive the gradient of the density dependent term with the second prescription. 
The projected gradient is given by an expression similar to eq.~(\ref{h20pdd}) but with $V_{DD}$
depending on the mixed density, eq.~(\ref{mixdens}), and the rearrangement term
 given by

\begin{eqnarray}
\delta {\breve{\Gamma}}_{kk'}^R ({\varphi}_{\pi},{\varphi}_{\nu}) & = &
\frac{ \langle \Phi | \displaystyle { \frac{ \partial {\hat V_{DD}} 
\left [ {\rho}_{{\varphi}_{\pi}{\varphi}_{\nu}} (\vec{r}) \right ] }
 {\partial {\rho}_{{\varphi}_{\pi}{\varphi}_{\nu}} (\vec{r}) } } f_{kk'}(\vec{r})
e^{i{\varphi}_{\pi} {\hat{N}}_{\pi}} 
e^{i{\varphi}_{\nu} {\hat{N}}_{\nu} } | \Phi \rangle } 
{\langle 
\Phi |  e^{i{\varphi}_{\pi} {\hat{N}}_{\pi}} 
e^{i{\varphi}_{\nu} {\hat{N}}_{\nu} } | \Phi \rangle }.
\end{eqnarray}
As before, we obtain
\begin{eqnarray}
\breve{\mathcal H}^{20,P}_{\mu \nu,\tau}  &  = & y^{\tau}_0 y^{\tau'}_0
 {\displaystyle Re}  \, \Biggl \{ \nonumber \\  
& & \sum_{l_{\tau}=0}^{[L/2]} \sum_{l_{\tau'}=0}^{[L/2]} {\mathcal C}_{l_{\tau}l_{\tau'}}
x_{l_{\tau}} x_{l_{\tau'}}
\left ( \breve{\mathcal H}^{20}_{\mu \nu , \tau} ({\varphi}_{l_{\tau}},
{\varphi}_{l_{\tau'}})
+ {\mathcal X}_{\mu \nu}({\varphi}_{l_{\tau}}) E^P_{DD}
({\varphi}_{l_{\tau}},{\varphi}_{l_{\tau'}}) \right )  + \nonumber \\
& + &   2   \,   
\sum_{l_{\tau}=1}^{[L/2]-1} \sum_{l_{\tau'}=1}^{[L/2]-1} 
x_{l_{\tau}} x^*_{l_{\tau'}}
\left ( \breve{\mathcal H}^{20}_{\mu \nu , \tau} ({\varphi}_{l_{\tau}},
{\varphi}_{l_{\tau'}}^*)
+ {\mathcal X}_{\mu \nu}({\varphi}_{l_{\tau}}) E^P_{DD}
({\varphi}_{l_{\tau}},{\varphi}_{l_{\tau'}}^*) \right ) \Biggr \}  \nonumber \\
& - & E^P_{DD} {\mathcal X}^{P\tau}_{\mu \nu}
\label{h20pmd}
\end{eqnarray}
where
\begin{eqnarray}
\breve{\mathcal H}^{20}_{\mu \nu , \tau} ({\varphi}_{l_{\tau}},
{\varphi}_{l_{\tau'}}) & = & 
{\left ( {\mathcal U}^{\dagger}({\varphi}_{l_{\tau}}) 
\left ( \breve{\Gamma}^{\tau\tau'} ({\varphi}_{l_{\tau}},
{\varphi}_{l_{\tau'}})  + \delta {\breve{\Gamma}}^R ({\varphi}_{l_{\tau}},
{\varphi}_{l_{\tau'}}) \right )  \right . }
{\mathcal V}^* ({\varphi}_{l_{\tau}}) - \nonumber \\
  & - & { \left . {\mathcal V}^{\dagger}({\varphi}_{l_{\tau}})
{ \left ( \breve{\Gamma}^{\tau \tau'} ({\varphi}_{l_{\tau}},
{\varphi}_{l_{\tau'}}) + \delta {\breve{\Gamma}}^R ({\varphi}_{l_{\tau}},
{\varphi}_{l_{\tau'}}) \right ) }^T  
{\mathcal U}^* ({\varphi}_{l_{\tau}})  \right ) }_{\mu \nu} \nonumber \\
E^P_{DD} ({\varphi}_{l_{\tau}},{\varphi}_{l_{\tau'}}) & = & 
\frac{1}{2} {\displaystyle Tr} \left ( \breve{\Gamma}^{\tau\tau'} ({\varphi}_{l_{\tau}},
{\varphi}_{l_{\tau'}}) \rho ({\varphi}_{l_{\tau}}) + 
 \breve{\Gamma}^{\tau\tau'} ({\varphi}_{l_{\tau}},
{\varphi}_{l_{\tau'}}) \rho ({\varphi}_{l_{\tau'}}) \right )
\label{hhdd} 
\end{eqnarray}
$\breve{\mathcal H}^{20}_{\mu \nu , \tau} ({\varphi}_{l_{\tau}},{\varphi}^*_{l_{\tau'}})$
and $E^P_{DD} ({\varphi}_{l_{\tau}},{\varphi}^*_{l_{\tau'}})$ in
eq.~(\ref{h20pmd}) are given 
by  the corresponding  expressions in eq.~(\ref{hhdd}) but with the replacement 
$ \rho ({\varphi}_{l_{\tau'})} \rightarrow (\rho ({\varphi}_{l_{\tau'}}))^*$.

\section{The Projection after variation method and the divergences problem.}
\label{sec:convL}

In this section we discuss some problems that may arise in PNP calculations
as well as some theoretical results.
As we have seen in eq.~(\ref{extwick}), the extended Wick theorem allows
to factorize matrix elements of the two-body part of the interaction 
in terms of the generalized matrix density and pairing tensors. The matrix
element of eq.~(\ref{extwick}) takes a simple form in the canonical basis. 
Using eqs.~(\ref{eq:rhophi}-\ref{eq:kap0phi}), we obtain

\begin{eqnarray}
\langle \Phi | & a^{\dagger}_{k_1} & a^{\dagger}_{k_2} a_{k_4} a_{k_3} |
\tilde{\Phi} \rangle  =  \langle \Phi | \tilde{\Phi} \rangle 
\left [ \rho_{k_3k_1}
\rho_{k_4k_2} -  \rho_{k_4k_1} \rho_{k_3k_2} + \kappa^{01}_{k_1k_2}
 \kappa^{10}_{k_3k_4} \right ] \nonumber \\
& = &  \langle \Phi | \tilde{\Phi} \rangle \left [ \frac{v_{k_1}^2 e^{2\img
\varphi}}{u_{k_1}^2 +v_{k_1}^2 e^{2 \img \varphi}} \cdot 
\frac{v_{k_2}^2 e^{2\img \varphi}}{u_{k_2}^2 +v_{k_2}^2 e^{2 \img \varphi}}
\left ( \delta_{k_3k_1} \delta_{k_2k_4} - \delta_{k_4k_1} \delta_{k_3k_2} 
\right ) \right. \nonumber \\
& + & \left . \frac{u_{k_1} v_{k_1}}{u_{k_1}^2 +v_{k_1}^2 e^{2 \img \varphi}} 
\cdot \frac{u_{k_3} v_{k_3} e^{2\img \varphi}}{u_{k_3}^2 +v_{k_3}^2 
e^{2 \img \varphi}}  \delta_{k_2 \bar{k}_1} \delta_{k_4 \bar{k_3}} \right ] 
\label{eq:overlap1}
\end{eqnarray}

The dangerous cases are for $\varphi=\pi/2$ and  $u_k^2=v_k^2$ when the
denominators become zero. 
Taking into account that $\langle \Phi | \tilde{\Phi} \rangle  = 
\prod_{k>0} (u_k^2 + v_k^2 e^{2 \img \varphi} ) $, it is easy to
see that  as long as there is only one pole 
or two poles but $k_1 \neq k_2$ (~or $k_3 \neq k_1$~)
there are no problems because they will be
cancel out by the multiplying overlap $\langle \Phi | \tilde{\Phi} \rangle$.
The only case where problems  arise is with the matrix elements of the form
$\langle \Phi | a^{\dagger}_{k} a^{\dagger}_{\bar{k}} a_{\bar{k}} a_{k} |
\tilde{\Phi} \rangle $, 
because in this case no cancellation with the norm
overlap $\langle \Phi | \tilde{\Phi} \rangle$ is possible.
It is easy to check, however, that in this case, though each term 
of eq.(\ref{eq:overlap1}) by itself is divergent, their sum gives a finite
contribution, namely \( v_{k}^2 \cdot e^{2 \img \varphi} \cdot 
\prod_{m>0,m \ne k} (u_m^2 + v_m^2 e^{2 \img \varphi} )\), which behaves
properly when $v_k^2 =  1/2$ and $\varphi=\pi/2$.
That means, as long as all
three terms of the Wick factorization of eq.~({\ref{extwick}}),
i.e., the Hartree, the Fock and the Bogoliubov term, are taken into account
no divergences appear in the PNP formalism. Notice that the arguments given
here are quite general and independent of the kind of force (Coulomb force
included) used in the calculations -see Appendix~\ref{appen:poles} for a 
detailed discussion. This point has already been noticed in ref.~\cite{Doe.98}. 
Obviously, one body operators will never cause problems to the PNP. 

In the case of the density dependent forces, one must be aware that this 
dependence itself
may cause problems. As we have seen before, two
different prescriptions have  mainly been used to treat the density dependence
of the Hamiltonian. As discussed in  Appendix~\ref{appen:poles}, the 
projected density prescription does not have divergences and the mixed density 
prescription
has an integrable divergence. For the moment, we use the
projected density prescription (PAV1), at the end of this section we show
 results with the mixed density prescription (PAV2).

Unfortunately most of the calculations performed so far with effective forces 
neglect exchange terms. Some of them  because different forces are used in the 
particle-hole and the particle-particle channel (like most of the Skyrme force
parametrizations and the  relativistic  mean field calculations).
The contributions to the p-p interaction from the force
used in the  p-h channel  being neglected
and vice-versa. Others, though with the same force in both channels,  like 
the Gogny force, 
neglect some exchange terms (Coulomb among others) just to save CPU time.
 Only recently HFB calculations including {\em all terms}  in a triaxial basis
has been performed with the Gogny force, see
\cite{AER.01}. 

Obviously, the  most interesting issue to address is to investigate how large 
the divergences are in the case that some exchange terms are neglected.
Though this point will ultimately depend on the particular force used in the
calculation, we shall illustrate it here for the Gogny force where we are able 
to perform the usual calculations which neglect some exchange terms and the
exact ones. To be precise, we present two types of calculations, the first one is 
the  usual  one used in calculations with this force in a triaxial basis
\cite{GG.83,ER.93}, namely, the pairing field 
is calculated {\it only} from  the Brink--Boeker term, i.e. all additional
pairing exchange terms are neglected and the Fock term of the Coulomb potential
is either neglected or calculated in an approximate way. We shall
call this approximation ${\rm HFB_s}$, $s$ for standard calculation
-see ref.~\cite{AER.01} for further details. The wave function solution of the
selfconsistent HFB equations with this approximate Hamiltonian will be denoted
$\Phi_s$. 
In the second calculation, the exact HFB equations (${\rm HFB_e}$) with the 
Gogny force are solved, i.e., {\it all} exchange terms 
are considered in the variational principle -see \cite{AER.01} for further
details. The wave function resulting from the
selfconsistent solution of the ${\rm HFB_e}$ equations will be denoted $\Phi_e$. 
 In a later stage the intrinsic wave functions obtained in these ways are
projected onto good particle number in order to calculate the projected energy.

In the calculations, we have expanded the quasiparticle operators in a triaxial 
harmonic oscillator basis. The configuration space was chosen by the condition
\begin{equation}
a_x n_x + a_y n_y + a_z n_z \leq N_0
\end{equation}
where the coefficients $a_i$ depend on the relations between the axes 
$q=R_z/R_x$ and $p=R_y/R_x$ in the form $a_x={(qp)}^{1/3}$,
$a_y=q^{1/3}p^{-2/3}$ and $a_z=p^{1/3}q^{-2/3}$. This basis has been 
symmetrized with respect to the simplex operation and is an eigenstate of the 
operator $\Pi_2 {\mathcal T}$, with $\Pi_2=\hat{P} e^{-i\pi\hat{J}_x}$, 
${\mathcal T}$ is the time--reversal operator,  and  $\hat{P}$ the parity 
operator~\cite{Goo79}. In the calculations we take $N_0 = 11.1$.
To take into account the rotational motion we add to the Hamiltonian the cranking 
term $-\omega \hat{J}_x$, as in ref.~\cite{ER.93,AER.01}. 

\begin{figure}[h]
\begin{center}
\parbox[c]{10cm}{\includegraphics[width=8cm]{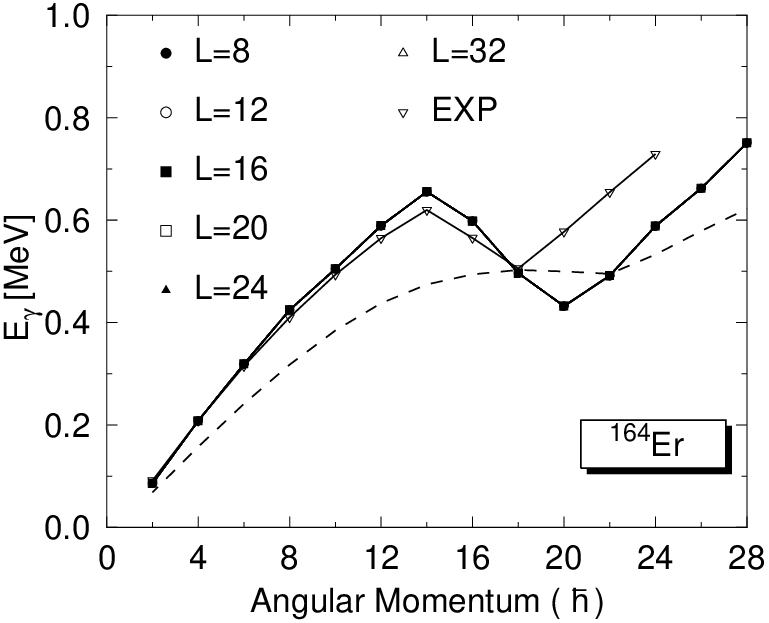}}
\end{center}
\caption{E$_{\gamma}$[MeV] versus angular momentum calculated
for different values of $L$ in the Fomenko expansion, together with the experimental values
and the  HFB$_e$ ones (dashed line). Notice that all points for different $L$ values lie
on top of each other for all angular momenta.} 
\label{fig:convLEX}
\end{figure}

 The most stringent test one can perform is to study the yrast band of a 
nucleus, \nuc{164}{Er} for example, since the  
pairing correlations  and  the quasiparticle
occupancies $v_k$ strongly depend  on the angular momentum. 
  To separate the different issues we shall first study  only the convergence 
  of the
projected energy in terms of the parameter $L$ of the Fomenko expansion. 
In order not to mix this issue with the divergence problem associated with 
the neglection of the exchange terms, we shall solve first the exact problem, 
i.e. the HFB$_e$ case. To illustrate more clearly our results we shall
perform particle number projection after the variation (PAV), i.e., first the
cranked HFB$_e$ equations with the Gogny force are solved, as in 
ref.~\cite{AER.01}.
This provides us with the wave function $\Phi_e$.
Then a projection on the particle number (PAV$_e$) is performed  onto
the wave function  $\Phi_e$ (for different  $L$ values )
to calculate the projected energy.
In fig.~\ref{fig:convLEX}, we display the results of the transition gamma ray
energy  in the PAV$_e$ approach for different $L$-values. We find that the 
points for the $L$-dependent calculation 
are on top of each other, which indicates a clean convergence and 
 that $L=8$ is already a good approximation to the exact PNP. 
This value is in agreement with earlier calculations \cite{Egi82,Carlo}. 
We find, by the way, a good agreement with the experiment \cite{ER164} up to 
$I=18 \hbar$ and a considerable improvement to the HFB$_e$ approach 
(dashed line). For higher spin values, the agreement with the experiment is 
not as good because the proton pairing  energy has already collapsed to zero.

\begin{figure}[h]
\begin{center}
\parbox[c]{10cm}{\includegraphics[width=8cm]{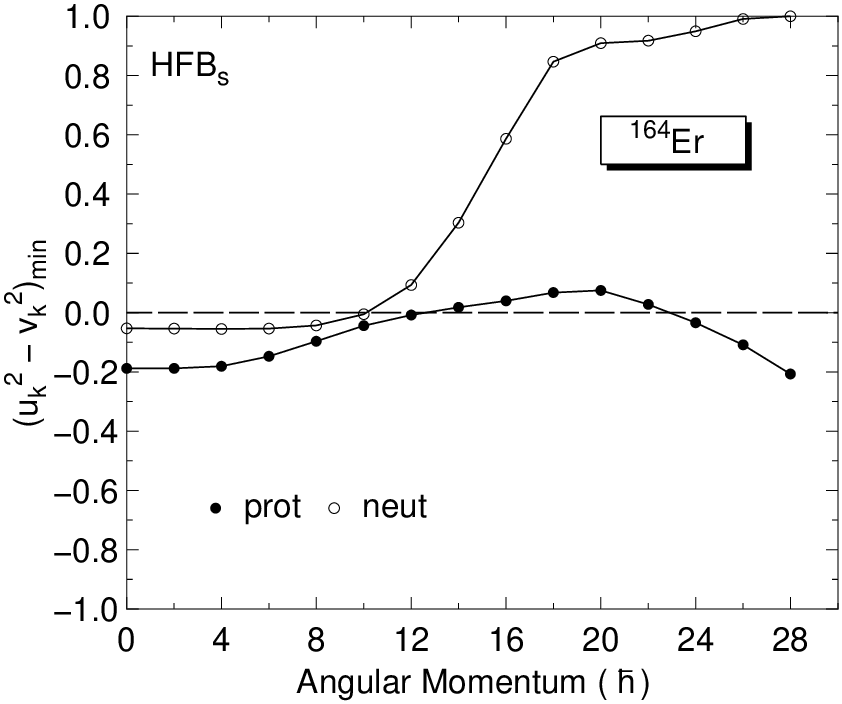}}
\end{center}
\caption{The smallest values of $(u_k^2 - u_k^2)$ along the Yrast line of \nuc{164}{Er}
for the HFB$_s$ in the canonical basis.}
\label{fig:uk2vk2}
\end{figure}

We now turn to the problem of the divergences in calculations where the exchange
terms are neglected. As we have seen in eq.~(\ref{eq:overlap1}) and
in Appendix \ref{appen:poles},  the poles
appear whenever $u_k^2 = v_k^2 =1/2$  and $\varphi_l= \pi/2$. The first 
condition, obviously, can not be eluded. In the non-selfconsistent PAV method 
because the occupancies $v_k^2$ of the wave function are already determined by 
the solution of the HFB equations and in the 
self-consistent VAP because  it would be against the variational principle. 
To investigate the behavior of the occupancies $v_k^2$, we have solved the 
selfconsistent cranked  HFB$_s$ equations
along the Yrast band for the nucleus  \nuc{164}{Er}. The wave functions 
$\Phi_s$ determined
in this way have been analyzed in the canonical basis.
In Fig.~\ref{fig:uk2vk2}, we show the smallest values of the quantity 
$u_k^2 - v_k^2 $, for protons
and neutrons, found in $\Phi_s$, as a function of the angular momentum.
 At small spins and
up to spin 8 $\hbar$, the quantity $u_k^2 - v_k^2 $ for neutrons 
is rather small (around $-0.05$), at spin 10 $\hbar$ it 
reaches the critical value $u_k^2 = v_k^2 $ and from this point on it steadily 
increases up to the maximal value at a spin value of 28 $\hbar$.
This behavior is typical of the high spin regime. At small angular momentum, 
the system is well paired and we may find a pair of nucleons close to the 
Fermi surface  with $u_k^2 \approx v_k^2 $. As the spin increases the pairs 
align (states start being predominantly  occupied or empty) and at high spins, 
when the pairing collapse takes place, 
the states become either fully occupied or empty.
For protons the situation is even worse, we find two critical points, one 
around $I= 12 \hbar$ and the other around $I= 22 \hbar$. 

 The second condition for having the pole, i.e. that $\varphi_l= \pi/2$,
is in principle avoidable, since $\varphi_l= \pi l/L$,  taking L odd 
in the Fomenko expansion should be enough. This recipe, as we shall 
see below, helps in some cases but it does not guarantees that one is error free.
The reason is very easy, to find convergence in the Fomenko expansion one must 
take  L large enough, but then one may get too close to the critical  $\pi/2$. 
That means  in this way one avoids to get nonsense but the results  may
not be reliable. 
As we have seen above and in Appendix~\ref{appen:poles}, in the neighborhood
of the poles we will get separately big contributions to the Hartree-Fock energy and to 
the pairing energy. If all exchange terms have been taken into account in the
calculations, the sum of the divergent terms provides a finite value. This 
cancellation, however, is not guaranteed if some exchange terms have been 
neglected, as in the ${\rm HFB_s}$ approach, which we now discuss. 
 In Table~\ref{tab:convL}, we show for the nucleus \nuc{164}{Er} the
differences in the pairing, eq.~(\ref{paienergy}), Hartree-Fock, eq.~(\ref{hfenergy}), and total, 
eq.~(\ref{eq:EP}), projected energies calculated for $L=8$ and $L=20$ 
(column labeled  L-EVEN) and for $L=9$ and $L=21$ (column labeled L-ODD) as a 
function of the angular momentum. 
In the even-L calculations in the PAV$_s$ approach, we find large  $\Delta E$
values in the pairing and the HF energies around $I=10, 12 \hbar $ and  
$I=24 \hbar $. These deviations do not cancel each other and give rise to large
$\Delta E_{tot}$, indicating that no convergence is reached. For odd-L values
the situation is much better. The largest $\Delta E_{tot}$ is around 20 keV
which is more than one order of magnitude smaller than in the even-L case.
Lastly, in the last column the difference of the total energy between $L=8$
and $L=9$ is displayed. We clearly observe that again large values
around $I=10, 12 \hbar $ and $I=24 \hbar $ are obtained. Interestingly, a comparison
with fig.~\ref{fig:uk2vk2} shows that these spin values are those where
$u_k^2 -v_k^2$ is very small, i.e. the poles. The neutron system is  
responsible for the increase at $I=10 \hbar $, see second column, and the 
proton one around $I=12 \hbar $ and $I=24 \hbar $.

\begin{table}[h]
\begin{center}
\begin{tabular}{c|ccc|ccc|c}
\hline
\hline
 &  \multicolumn{3}{c|}{L-EVEN}&\multicolumn{3}{c|}{L-ODD} &\\ 
\hline
I($\hbar$) & $\Delta E_{pai}$ & $\Delta E_{hf}$ & 
$\Delta E_{tot}$  & $\Delta E_{pai}$ & $\Delta E_{hf}$ & 
$\Delta E_{tot}$  & $ E_{tot}^8- E_{tot}^9$\\
\hline
 0  &	-1.12  &  -2.45 &   -3.55&    1.46  &  0.94 &  2.39  & 5.8 \\
 2  &	-1.21  &  -2.43 &   -3.64&    1.50  &  0.94 &  2.44  & 5.9 \\
 4  &	-1.71  &  -2.61 &   -4.31&    1.86  &  1.02 &  2.88  & 7.0 \\
 6  &	-4.19  &  -3.62 &   -7.81&    3.75  &  1.36 &  5.11  & 13.1 \\
 8  &  -11.76  &  -6.99 &  -18.76&    8.55  &  1.80 & 10.35  & 31.6 \\
10  &	57.62  & -99.99 &  -42.37&    12.67 &  0.31 & 12.98  & 71.0 \\
12  & -347.66  &  22.27 & -325.39&    8.40  & -4.64 &  3.76  & 542.4 \\
14  &  149.53  & -10.35 &  139.17&   -7.09  &  0.27 & -6.81  & -232.1 \\
16  &	61.24  &  -6.07 &   55.17&  -11.96  &  1.19 &-10.77  & -92.2 \\
18  &	27.70  &  -3.33 &   24.37&  -10.88  &  1.31 & -9.57  & -41.0 \\
20  &	24.19  &  -3.12 &   21.07&  -10.55  &  1.36 & -9.19  & -35.5 \\
22  &  124.15  & -15.49 &  108.66&  -13.28  &  1.66 &-11.62  & -181.4\\
24  & -144.09  &  17.41 & -126.68&   21.92  & -2.65 & 19.27  & 211.8 \\
26  &  -38.19  &   4.45 &  -33.75&   21.73  & -2.53 & 19.20  & 56.9 \\
28  &	-9.22  &   1.01 &   -8.21&    5.71  & -0.63 &  5.09  & 13.4 \\
\hline
\hline
\end{tabular}
\caption{Differences (in keV) between the projected energies calculated with 
$L=9$ ($E_i^9$) and $L=21$ ($E_i^{21}$) and $L=8$ ($E_i^8$) and $L=20$ ($E_i^{20}$) 
in the Fomenko expansion for the nucleus \nuc{164}{Er}, calculated in the PAV$_s$
approach. The last column displays the difference between the total energy
calculated with $L=8$ and $L=9$.}
\label{tab:convL}
\end{center}
\end{table}

\begin{figure}[h]
\begin{center}
\parbox[c]{14cm}{\includegraphics[width=12cm]{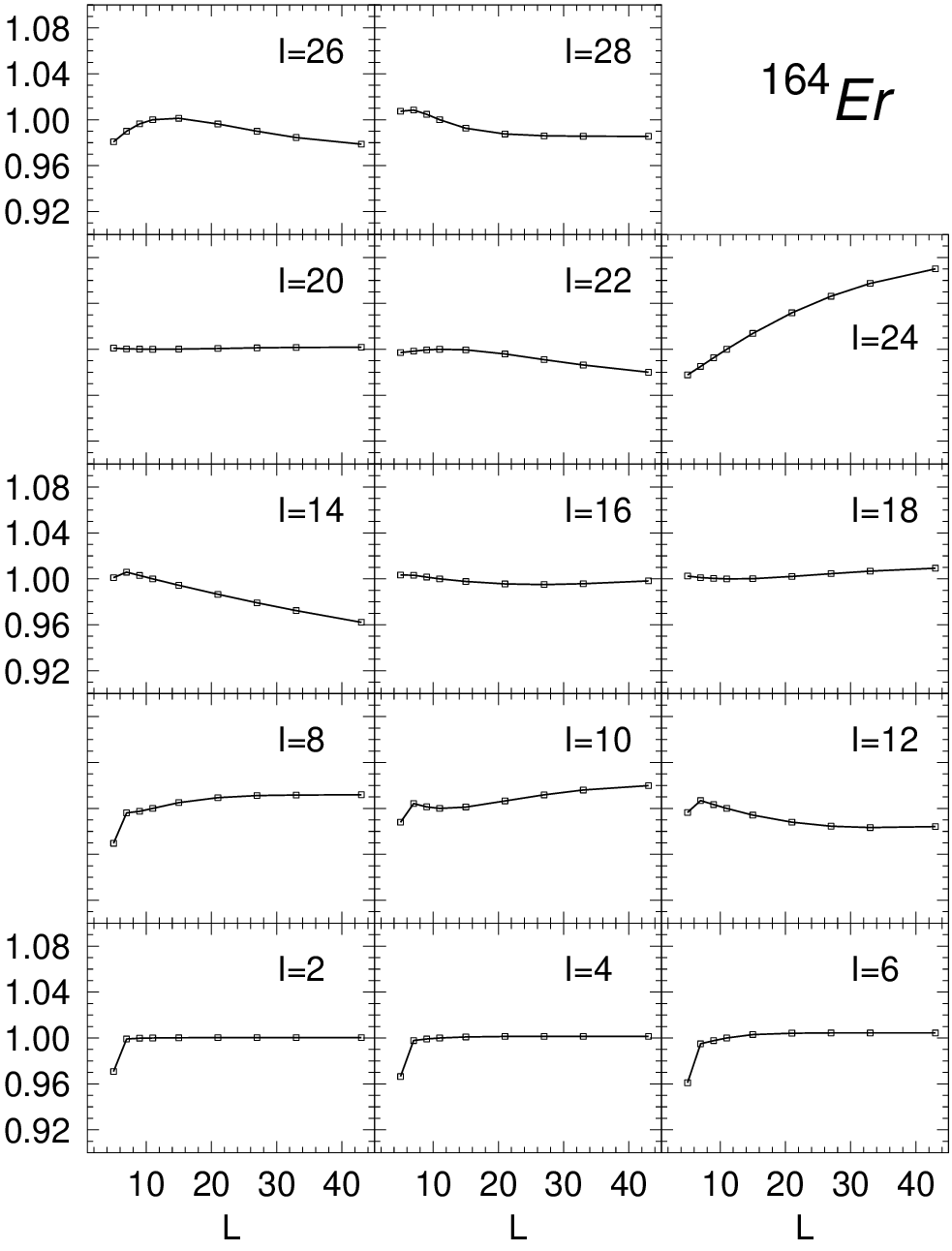}}
\end{center}
\caption{The transition energies E$_{\gamma}$[MeV] for a given  angular momentum
as a function of the parameter $L$ of the Fomenko expansion.}
\label{fig:Lconverg}
\end{figure}

>From this table, we conclude that even-L values should be avoided 
and that the total energy for odd-L is a function of $L$. Usually, we are
not interested in total energies but in relative ones, for instance in the
$\gamma$-ray energy. Now the question  is whether we can find a kind of plateau 
condition that provides us the optimal L-value for all $I$-values.
 In Fig.~\ref{fig:Lconverg} we display the quantity
$E_\gamma^L (I)/ E_\gamma^{11}(I)$  as a function of $L$. 
We have chosen the value $L=11$ to normalize the  $\gamma$-ray energies 
because it is large enough that in the absence of poles
a good convergence would be found, and small enough that in the presence
of poles we do not come too close to the critical $\varphi_L  = \pi /2$.
We found a  good plateau for spin values 2 to $6 \hbar $ and  a rather
good one for $I= 16, 18$ and $20 \hbar $, the worst cases  are for 
 $I=14 \hbar$ and  $I=24 \hbar$. This behavior is again strongly correlated 
with the  nodes in Fig.~\ref{fig:uk2vk2}. We observe that the largest 
deviations of the {\em most reasonable} values are about 7 \%. It is
clear that more sensitive quantities like second moments of inertia will
show larger uncertainties.

All calculations of projected energies shown up to this point have been done 
with the prescription of eq.~(\ref{eq:EPDD1}) for the density dependent part
of the force, namely the projected density.  

\begin{table}[h]
\begin{center}
\begin{tabular}{c|c|cc|cc|c}
\hline
\multicolumn{2}{c|}{}&  \multicolumn{2}{c|}{DD1}&\multicolumn{3}{c}{DD2} \\
\hline
I($\hbar$) & $\Delta E^{8-20}_{pai}$ & 
$\Delta E^{8-20}_{hf}$   & $\Delta E^{8-20}_{tot}$  &
$\Delta E^{8-20}_{hf}$  
 & $\Delta E^{8-20}_{tot}$  &  $\Delta E^{9-21}_{tot}$   \\
\hline
00 &  -17.27  &   17.28   &  0.01     &  -6.03     & -23.30 &  -0.56 \\
02 &  -26.45  &   26.45   &  0.01     &  -6.60     & -33.05 &  -0.42 \\
04 &  145.93  &  -145.92  &  0.01     & -95.88     &  50.05 &	0.08 \\
06 &   13.91  &   -13.90  &  0.01     &  18.93     &  32.85 &	1.21 \\
08 & 	6.24  &    -6.23  &  0.01     &   5.82     &  12.05 &	3.17 \\
10 & 	3.11  &    -3.11  &  0        &   2.61     &  5.72  &	3.16 \\
12 & 	1.15  &    -1.15  &  0        &   0.98     &  2.13  &	1.27 \\
14 & 	0.36  &    -0.36  &  0        &   0.32     &  0.68  &	0.38 \\
16 & 	0.12  &    -0.12  &  0        &   0.10     &  0.22  &	0.10 \\
18 & 	0.02  &    -0.02  &  0        &   0.02     &  0.04  &	0.01 \\
20 & 	0.01  &    -0.01  &  0        &   0.01     &  0.02  &	   0  \\
22 &       0  &	       0  &  0        &      0     &  0.01  &	   0  \\
24 &       0  &	       0  &  0        &      0     &  0     &	   0 \\
26 &       0  &	       0  &  0        &      0     &  0     &	   0 \\
28 &       0  &	       0  &  0        &      0     &  0     &	   0 \\
\hline
\end{tabular}
\caption{Differences $\Delta E$ (in keV) between the projected energies 
calculated with $L=8$ ($E_i^8$) and $L=20$ ($E_i^{20}$) in the Fomenko 
expansion for the nucleus \nuc{164}{Er} taking into account all the 
contributions of the force to the fields. [DD1] and [DD2] are meaning the
prescription used for the density--dependent term. In the last column 
$\Delta E_{tot}$ is the difference between the projected energies calculated 
with  $L=9$ and $L=21$.}
\label{tab:convLEX}
\end{center}
\end{table}
  We shall now discuss the convergence of our results with respect to the
density dependence of the Hamiltonian used in the calculations. In order
to isolate this issue, we do not neglect any exchange terms in the calculations,
that means we solve the HFB$_e$ equations. This provides us with the wave function
$\Phi_e$ which we use to calculate the projected energy, which can calculated
according to two prescriptions. Notice that $\Phi_e$ is different from $\Phi_s$
and that the levels occupancies of both wave functions are not necessarily the 
same. As it is shown in section \ref{sect:approden} the prescription 
using the projected density, which we shall label DD1, was divergence free, 
while the second prescription, the mixed density one 
(section \ref{sect:apmixden}),  from now on called DD2, was 
showing an integrable divergence. As in Table~\ref{tab:convL} we shall
calculate separately  the pairing and the Hartree-Fock 
contributions to the total energy.
Since the density--dependent term does not contribute to the pairing energy 
 (the parameter $x_0$ in the Gogny force is equal to $1$) this energy is 
independent of the prescription used for the density dependent term.
In Table~\ref{tab:convLEX} we display  the differences $\Delta E^{8-20}$ 
  of projected 
energies calculated with the pairing energy, eq.~(\ref{paienergy}),
the  HF energy, eq.~(\ref{hfenergy}) and the sum of both, the total energy.
 At low and medium spins, the difference of  pairing energies for $L=8$ and 
 20 is not equal to zero indicating the known divergences.
 We find the largest value for $I= 4 \hbar$ indicating 
that the wave function $\Phi_e$ has  a level occupancy at this spin
value such that $u_k^2 \approx v_k^2$. Inspection of the canonical basis
indicates that the orbital is  a neutron one, the proton system in this case does
not have orbitals with  $v_k^2$ close to 1/2.
 At high spins, the wave function
$\Phi_e$ is not paired in the proton system  (see ref.~\cite{AER.01}) 
and in the neutron system it does not have orbitals with  $v_k^2$ close to 1/2.
 To get convergence,  either
with the DD1 or the DD2 prescription, the difference of the total energy 
calculated with different L values, must be zero.
This is the case in the DD1 approach where, even with even-L values,
the HF energy differences are almost exactly the same as the pairing ones. 
As a result, both cancel out and the total energy is convergent, as expected.
With the DD2 prescription and even-L values, however,  the HF results are
different from the pairing ones and as a result no convergence for the
total energy is found. We know, however, that the DD2 divergence is integrable,
that means if the integral is carefully performed the convergence must be found.
This is what happens if we do the calculations with odd-L values (see last
column) where the largest deviation found is 3 keV. Obviously, more sophisticated
integration methods will provide better convergence. 

\begin{figure}[h]
\begin{center}
\parbox[c]{10cm}{\includegraphics[width=8cm]{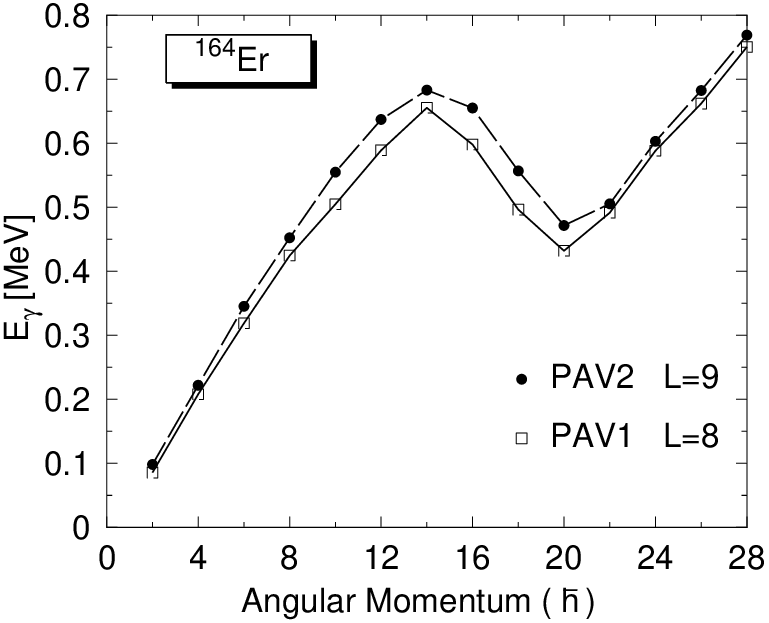}}
\end{center}
\caption{E$_{\gamma}$[MeV] versus angular momentum for both prescriptions of the
density.}
\label{fig:dd1dd2}
\end{figure}

An interesting point concerning the density prescriptions DD1 and DD2
is to know if there is a big difference in the results calculated with
DD1 or DD2.  In Fig.~\ref{fig:dd1dd2} we show the $\gamma$-ray energies
along the yrast line of the nucleus $^{164}$Er in the HFB$_{\rm e}$ with
the DD1 (in the figure PAV1) and with DD2 (in the figure PAV2). As we can see
the behavior is very similar and  the physics is clearly the same. A similar 
conclusion was found in \cite{VER.97} in the Lipkin-Nogami approach. 

\section{ The variation after projection method~: The nuclei $^{48}$Cr and $^{50}$Cr}
\label{sec:VAPR}
In ref. \cite{CEM.95} and \cite{MPR.96}  a good  agreement between
the HFB approximation, the exact shell model diagonalization and the 
experimental results was found for the rotational states of the nuclei 
 $^{48}$Cr and $^{50}$Cr. Only the $\gamma$-ray energies 
along the yrast bands in the HFB approach  were systematically 
around 0.5 MeV smaller than experimentally observed. 
A simulation with the exact shell model diagonalization
in \cite{CEM.95} showed that the shift was due to a deficient treatment
of the pairing correlations of the HFB method in the weak pairing regime.
  It is well known that the variation after projection method
is able to provide pairing correlations even at the limiting case 
when the HFB approach collapses to the HF one.
In this section we shall investigate the mentioned nuclei in the VAP method.

\begin{figure}[h]
\begin{center}
\parbox[c]{16cm}{\includegraphics[width=14cm]{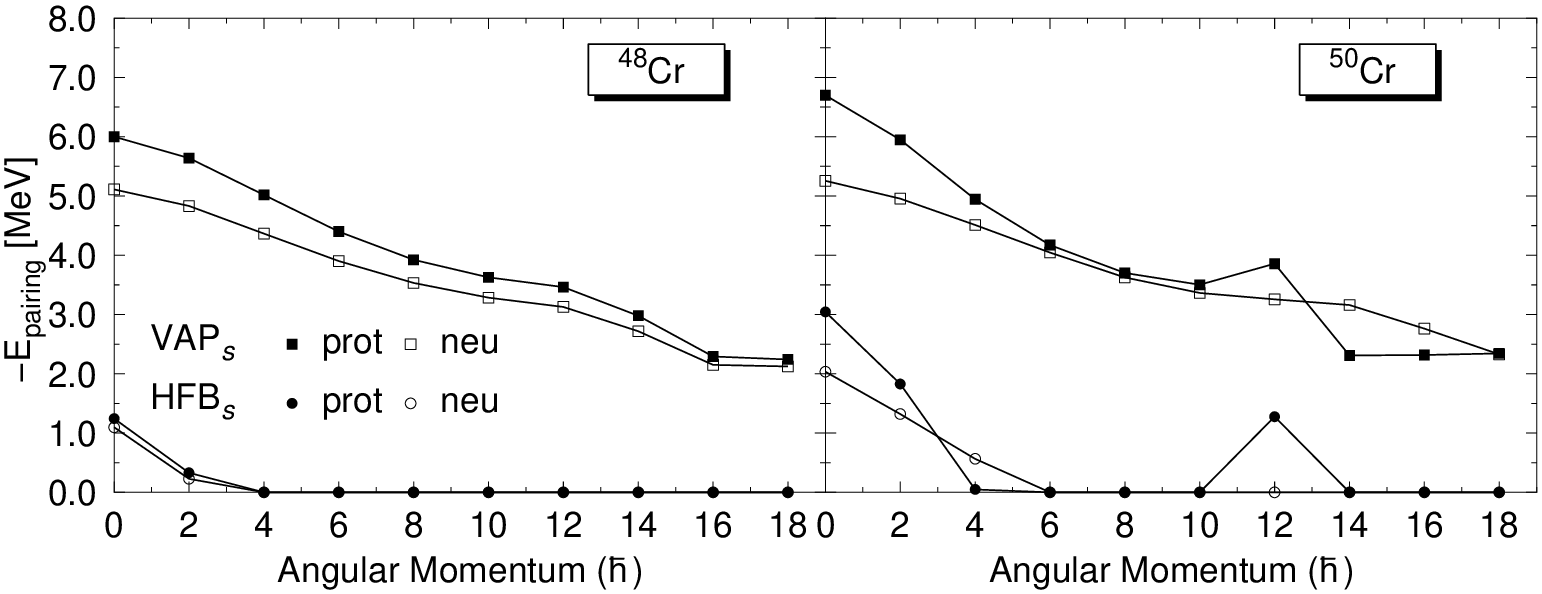}}
\end{center}
\caption{Pairing correlation energies in the HFB$_s$ and VAP$_s$ approaches
for the nucleus $^{48}$Cr and $^{50}$Cr respectively.}
\label{fig:paircrom_s}
\end{figure}

 The HFB calculations of refs. \cite{CEM.95} and \cite{MPR.96} were performed
in the standard approximation, i.e. HFB$_s$. For nuclei like these with little 
pairing -low level density at the Fermi surface- is very unlikely to find 
orbitals with $v_k^2 = u_k^2$, nevertheless we have  checked during
the minimization process of the VAP$_s$ that no orbitals were populated in that 
way. Additionally, we have performed also the HFB$_e$ and the VAP$_e$ 
calculations.
In Fig.~\ref{fig:paircrom_s}, we represent the HFB$_s$  and VAP$_s$ 
(see eq.(\ref{paienergy})) pairing energies for the nuclei $^{48}$Cr, left hand side,
 and $^{50}$Cr, right hand side.   In the HFB$_s$ approach the pairing
energies are rather small, for $^{48}$Cr practically zero for all spin values,
with the exception of the $I=0 $ and $ 2 \hbar$ values where they are
very small.
For $^{50}$Cr they are also very small and only for $I=0, 2, 4$ and $ 12 \hbar$ 
 differ from zero.  In both nuclei we find the Coriolis antipairing 
effect that leads to
a sharp transition from a (weakly) superfluid system to a normal one. The
VAP$_s$ solutions, on the other hand, are well paired and no sharp transition
 is seen, only a smooth decrease in the correlation energy is found.  

\begin{figure}[h]
\begin{center}
\parbox[c]{16cm}{\includegraphics[width=14cm]{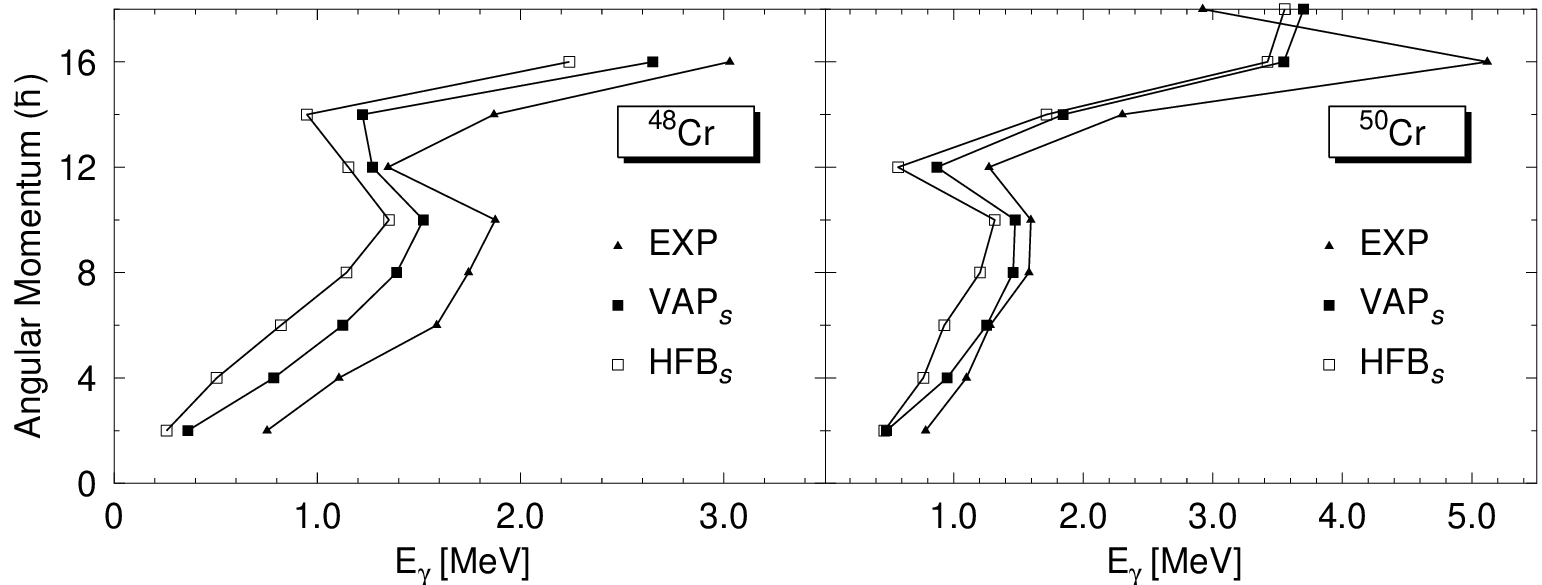}}
\end{center}
\caption{ The transition ${\gamma}$-ray energies versus the angular momentum
in the HFB$_s$ and VAP$_s$ approaches for the nucleus $^{48}$Cr
and $^{50}$Cr. The experimental results are also shown.}
\label{fig:gammacrom_s}
\end{figure}

In Fig.~\ref{fig:gammacrom_s}, the ${\gamma}$-ray energies along the yrast
band are displayed in the HFB$_s$ and VAP$_s$ approaches together with
the experimental results for $^{48}$Cr, \cite{exp.C48} and $^{50}$Cr,
\cite{exp.C50}. As discussed in \cite{CEM.95}, the 
HFB$_s$ ${\gamma}$-ray energies for $^{48}$Cr show the same trend as 
the experimental ones 
but  are about 0.5 MeV smaller. The VAP$_s$ approach, on the other hand,
due to the larger pairing energy leads to smaller moment of inertia 
improving the agreement with the experiment. The  VAP$_s$ results come closer 
to the
experiment and the backbending is slightly better described. For  $^{50}$Cr 
the situation is similar, the VAP$_s$ approach improves considerably 
the HFB$_s$
 one specially at low and medium spins. 
One may ask, however, why we do not get enough pairing correlations in the
VAP$_s$ method as to match the experimental results. There are several possible
answers, first in our calculations we do not include p-n pairing which is
known \cite{T=0pair} to play an important role in these nuclei. Second,
the fact that in these nuclei, in the HFB approximation, there are barely any  
pairing correlations indicates that the potential energy curve is rather
flat against pairing fluctuations. This kind  of fluctuations is not included 
in the PNP method.         

\begin{figure}[h]
\begin{center}
\parbox[c]{16cm}{\includegraphics[width=14cm]{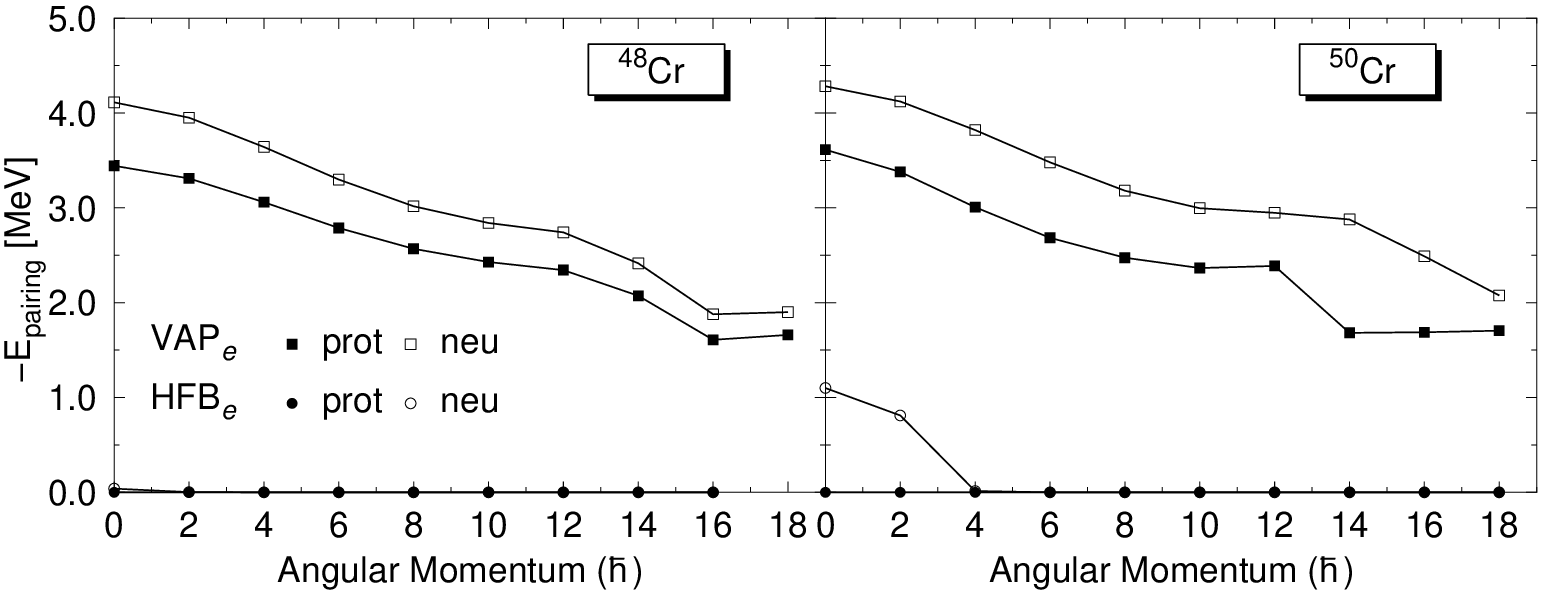}}
\end{center}
\caption{Pairing correlation energies in the HFB$_e$ and VAP$_e$ approaches
for the nuclei $^{48}$Cr and $^{50}$Cr respectively.}
\label{fig:paircrom_e}
\end{figure}

\begin{figure}[h]
\begin{center}
\parbox[c]{16cm}{\includegraphics[width=14cm]{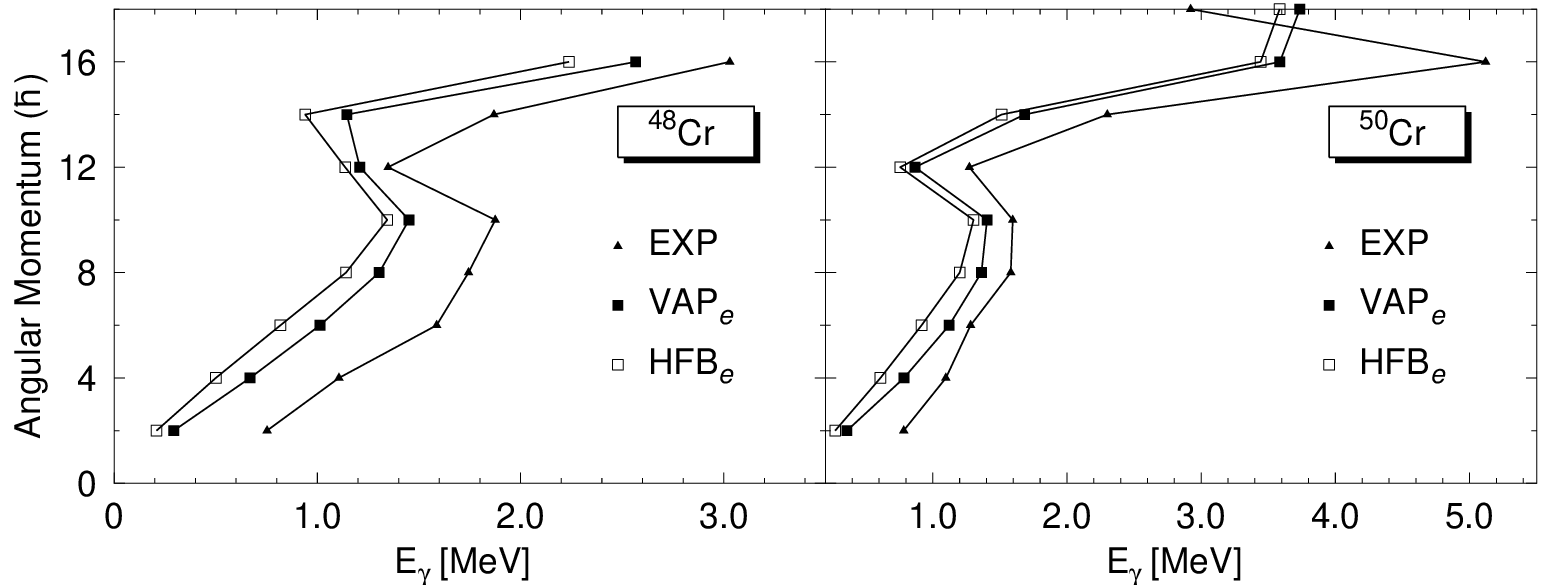}}
\end{center}
\caption{The transition ${\gamma}$-ray energies versus the angular momentum
in the HFB$_e$ and VAP$_e$ approaches for the nuclei $^{48}$Cr 
and $^{50}$Cr. The experimental results are also shown.}
\label{fig:gammacrom_e}
\end{figure}

In Figs.~\ref{fig:paircrom_e} and \ref{fig:gammacrom_e}, we finally present
the results of the exact HFB calculations,  HFB$_e$, and the corresponding
VAP$_e$ method. The pairing correlation energies are represented in 
Fig.~\ref{fig:paircrom_e}. As expected \cite{AER.01}, the neglected pairing
terms in the HFB$_s$ approach have an antipairing effect, mainly
the Coulomb term.  The small pairing energy of the HFB$_s$ approach
is quenched in the HFB$_e$, in particular the proton pairing energies
completely vanish. The pairing energies in the  VAP$_e$ approach display
 this effect even more clearly. The total VAP$_e$ pairing energy is 
 considerably reduced with respect to the  VAP$_s$ approach.
  Furthermore, while  in the VAP$_s$ the proton pairing energy  is always 
larger than the neutron one in the VAP$_e$ the proton pairing energy becomes
always smaller than the neutron one.

 In Fig.~\ref{fig:gammacrom_e} we finally present the $\gamma$-ray energies
in the HFB$_e$ and the VAP$_e$ together with the experimental ones.  Since
the pairing energies are negligible in the  HFB$_e$ and HFB$_s$ for both 
nuclei, both sets of $\gamma$-ray energies practically coincide. With 
respect to the VAP$_e$ values they are of the same quality as the VAP$_s$
though somewhat smaller than these one, as expected from the behavior of the
pairing correlations.

\section{Conclusions.}

In this paper we have derived the expressions needed to perform
a particle number projected HFB calculation  with  a finite--range and 
density dependent force in the triaxial basis. We have studied and
thoroughly discussed the divergences that appear for special values
of orbital occupancies when exchange terms are neglected. 
We have performed exact and approximate calculations (neglecting 
some exchange terms) and found that in the latter case some problems 
associated with the convergence of the particle number projection arise
leading in some cases to  an inaccuracy of  about 10 \% in the values
of the $\gamma$-ray energies. Interestingly, {\em all terms} of the force,
even of the Coulomb part, must be taken into account. This feature presents
a challenge for interactions where a given force is used in the 
particle-hole channel and a different one in the particle-particle channel. 

  We have furthermore performed variation after the projection for the 
$^{48,50}$Cr nuclei with the Gogny force. We found strong pairing correlations, 
at variance with the weekly correlated HFB approach. These correlations smoothly 
decrease with increasing angular momentum indicating that no sharp phase
transition takes place as predicted by the HFB  approach. The projected
calculations improve the HFB ones but some correlations (probably T=0 pairing) 
not present in our approach are still missing.

%\section{Acknowledgements}
\begin{ack}
This work was supported in part by DGICyT, Spain under Project PB97-0023.
One of us (M.A) acknowledges a grant from the Spanish Ministry of Education
under Project PB94-0164.
\end{ack}

%\newpage

\appendix

\section{Appendix A: The Gogny Force.}

\label{appen:GogF}

In the calculations we use  the Gogny interaction
 \cite{Go.75} as the effective force .
The main ingredients of this force are the phenomenological density dependent
term which was introduced to simulate the effect of a G--matrix interaction and
the finite range of the force which allows to obtain the Pairing and 
Hartree--Fock fields from the same interaction.
We use the parametrization D1S, which was fixed by 
Berger {\em et al.}~\cite{Ber91}. The force is given by
\begin{eqnarray}
v_{12} & = & \sum_{i=1}^2 e^{-{(\vec{r}_1 - \vec{r}_2)}^2/\mu_i^2} (W_i +
B_iP_{\sigma} -H_i P_{\tau} -M_i P_{\sigma} P_{\tau} ) + \nonumber \\
& + & W_{LS} (\vec{\sigma}_1 + \vec{\sigma}_2) \vec{k} \times \delta(\vec{r}_1 -
\vec{r}_2) \vec{k} + \nonumber \\
& + & t_3(1+x_0 P_{\sigma}) \delta (\vec{r}_1 -\vec{r}_2) \rho^{1/3} \left (
\frac{1}{2} (\vec{r}_1 + \vec{r}_2 )\right ),
\label{eq:vgog}
\end{eqnarray}
and the Coulomb force
\begin{equation}
v_{12}^C = (1+2\tau_{1z})(1+2\tau_{2z}) \frac{e^2} {|\vec{r}_1-\vec{r}_2 | }.
\end{equation}
The density operator, $\hat{\rho} (\vec{r})$, is given by 
\begin{equation}
\hat{\rho} (\vec{r}) =  \sum_{i=1}^{A} \delta (\vec{r} -\vec{r}_i)  =
 \sum_{ij} \phi_i^* (\vec{r}) \phi_j (\vec{r}) \langle S_i | S_j \rangle
c_i^{\dagger}c_j = \sum_{ij} f_{ij} (\vec{r}) c_i^{\dagger} c_j \, .
\label{rhodef}
\end{equation}
In the two-body interaction used in the calculations we also include 
 the one--body and two--body  center  of mass corrections.
\begin{equation}
\hat{T}  =  \sum_i \frac{{\vec{p}_i}^{\,2}}{2m} \left ( 1 -\frac{1}{A} \right ) -
\frac{1}{Am} \sum_{i>j} \, \vec{p}_i \cdot \vec{p}_j
\end{equation}

\section{Appendix B: Calculation of the projected energy in the canonical 
basis.}
\label{appen:poles}
In this appendix we  show that the projected energy is always well--defined 
if the direct (Hartree), exchange (Fock) and pairing 
terms of each component of the force are taken into account, i.e. if no
exchange terms are neglected.

 The projected energy can be written
in the following way~\footnote{To simplify the expression we consider the
projection onto good particle number only in one dimension in the gauge space.}
\begin{equation}
E^P  =   Tr (t \rho^P) + E^P_{hf} + E^P_{g}
\label{eq:EP}
\end{equation}
where
\begin{eqnarray}
E^P_{hf} & = & \frac{1}{2} \sum_{k_1k_2k_3k_4}
\bar{v}_{k_1k_2k_3k_4} 
\sum_l y_l \, \rho_{k_3k_1}(\varphi_l) \rho_{k_4k_2}(\varphi_l) 
\label{eq:EPHF}\\
E^P_{g} & = & \frac{1}{4} \sum_{k_1k_2k_3k_4} \bar{v}_{k_1k_2k_3k_4} 
\sum_l y_l \, \kappa^{01}_{k_1k_2}(\varphi_l) \kappa^{10}_{k_3k_4}(\varphi_l)
\label{eq:EPPA}
\end{eqnarray}

These energies are given in
terms of the $\varphi_l$-dependent  density matrix $\rho(\varphi_l)$
and the  pairing tensors $\kappa^{01}(\varphi_l)$
and $\kappa^{10}(\varphi_l)$. These quantities can be easily
evaluated in the canonical basis.
In this basis, denoted by $\{a,a^{\dagger} \}$, the usual density
matrix $\rho_{{k_1}{k_2}}  = \langle \Phi | a_{k_2}^{\dagger} a_{k_1} | \Phi \rangle$  is
diagonal and the usual pairing tensor
 $\kappa_{{k_1}{k_2}} = \langle \Phi | a_{k_2} a_{k_1} | \Phi \rangle$ 
has the canonical form. In this basis the quantities we are interested in
are given by
\begin{eqnarray}
\rho_{k_1k_2} (\varphi_l) & = & \frac{\langle \Phi | a_{k_2}^{\dagger} a_{k_1}
 e^{\img \varphi_l \hat{N}} | \Phi \rangle }{ \langle \Phi | 
e^{\img \varphi_l \hat{N}} | \Phi  \rangle } \nonumber \\
\kappa^{01}_{k_1k_2} (\varphi_l) & = &  \frac{\langle \Phi | a_{k_1}^{\dagger} 
a_{k_2}^{\dagger} e^{\img \varphi_l \hat{N}} | \Phi \rangle }{ \langle \Phi | 
e^{\img \varphi_l \hat{N}} | \Phi \rangle }, \quad  \quad
\kappa^{10}_{k_1k_2} (\varphi_l) = \frac{\langle \Phi | a_{k_2} a_{k_1} 
e^{\img \varphi_l \hat{N}} | \Phi \rangle }{ \langle \Phi | 
e^{\img \varphi_l \hat{N}} | \Phi \rangle }.
\end{eqnarray}
These matrix elements can be easily evaluated
\begin{eqnarray}
\langle \Phi | e^{\img \varphi_l \hat{N}} | \Phi \rangle & = & \prod_{k>0} 
\langle - | (u_k + v_k a_{\bar{k}}a_k) e^{\img \varphi_l \hat{N}} 
(u_k + v_k a_k^{\dagger}a_{\bar{k}}^{\dagger}) |- \rangle  \nonumber \\
& = &  \prod_{k>0} \langle - | (u_k + v_k a_{\bar{k}}a_k)
(u_k + v_k e^{2 \img \varphi_l} a_k^{\dagger}a_{\bar{k}}^{\dagger})  |- \rangle
\\ \nonumber
& = &
\prod_{k>0} (u_k^2 + v_k^2 e^{2 \img \varphi_l} ) | - \rangle,
\end{eqnarray}
here $\bar{k}$ denotes the canonical conjugated state of $k$.
To calculate $\langle \Phi | a_{k_2}^{\dagger} a_{k_1}
e^{\img \varphi_l \hat{N}} | \Phi \rangle $, we shall proceed in the same way.
 Let us first assume $k_2 \ne k_1$, then
\begin{equation}
\langle \Phi | a_{k_2}^{\dagger} a_{k_1} e^{\img \varphi_l \hat{N}} | \Phi \rangle  = 
\prod_{k>0} \langle - | (u_k + v_k a_{\bar{k}} a_k) a_{k_2}^{\dagger} a_{k_1}
(u_k + v_k e^{2 \img \varphi_l} a_k^{\dagger}a_{\bar{k}}^{\dagger})  |- \rangle =
0,
\end{equation}
that means, the $\varphi_l$-dependent density matrix $\rho_{{k_1}{k_2}} (\varphi_l)$, like 
the normal density matrix $\rho_{{k_1}{k_2}}$, is diagonal in the canonical basis. 
 For $k_2=k_1=k$ we obtain
\begin{eqnarray}
\langle \Phi | a_k^{\dagger} a_k e^{\img \varphi_l \hat{N}} | \Phi \rangle = 
v_k^2 \cdot e^{2 \img \varphi_l} \cdot \prod_{m>0,m \ne k} 
(u_m^2 + v_m^2 e^{2 \img \varphi_l} ).
\end{eqnarray}
Therefore, in the canonical basis, the matrix $\rho_{km} (\varphi_l)$ is given
by
\begin{eqnarray}
\rho_{km} (\varphi_l) = \delta_{km} \cdot \frac{v_k^2 \cdot e^{2 \img \varphi_l} } 
{u_k^2 + v_k^2 e^{2 \img \varphi_l} }. 
\end{eqnarray}
We see in this expression that for $u_k = v_k$ and $\varphi_l=\pi /2$, the
matrix element diverges and one has to be very careful in the calculations.
It is easy to show that
\(\rho_{\bar{k} \bar{m}}(\varphi_l) = \rho_{km} (\varphi_l)\) .

Let us now calculate the $\varphi_l$-dependent pairing tensors in the canonical
 basis
\begin{eqnarray}
\langle \Phi | a_{k_2}^{\dagger} a_{k_1}^{\dagger} e^{\img \hat{N} \varphi_l} |
\Phi \rangle &  = &  
\prod_{m >0 } \langle - | (u_m + v_m a_{\bar{m}} a_m ) 
a_{k_2}^{\dagger} a_{k_1}^{\dagger} (u_m + v_m a_m^{\dagger} a_{\bar{m}}^{\dagger} 
e^{2 \img \varphi_l} | - \rangle 
\nonumber \\
& = & \delta_{k_1 \bar{k}_2} u_{k_2} v_{k_2}  \prod_{m > 0, m\neq k_2} (u_m^2 + v_m^2 
\cdot e^{2 \img \varphi_l} ) 
\end{eqnarray}
in the same way
\begin{eqnarray}
\langle \Phi | a_{k_1} a_{k_2} e^{\img \varphi_l \hat{N}} | \Phi \rangle = 
\delta_{k_1 \bar{k}_2} u_{k_2} v_{k_2}  \prod_{m>0,m \neq k_2} (u_m^2 + v_m^2 \cdot 
e^{2 \img \varphi_l} ). 
\end{eqnarray}
The non-zero matrix elements are given by
\begin{eqnarray}
 \kappa_{k\bar{k}}^{01} = \frac{u_k v_k}{u_k^2 + v_k^2 \, e^{2 \img \varphi_l} }, 
\quad   \quad \quad  \kappa_{k \bar{k}}^{10} = \frac{u_k v_k e^{2 \img \varphi_l} } 
{u_k^2 + v_k^2 \, e^{2 \img \varphi_l} }
\end{eqnarray}
obviously, $\kappa_{k \bar{k}}^{01} = - \kappa_{\bar{k}k}^{01}$ and 
 $\kappa_{k\bar{k}}^{10} = - \kappa_{\bar{k}k}^{10}$. Again the 
dangerous terms are for $u_k = v_k$ and $\varphi_L = \pi /2$.

>From eq.~(\ref{eq:EPHF}) and using the expressions derived above, we obtain
\begin{eqnarray}
E_{hf}^P & =  & 
\frac{1}{2} \sum_{k_1 k_2 k_3 k_4} {\bar{v}}_{k_1 k_2 k_3 k_4} 
\sum_l y_l \rho_{kk}(\varphi_l) \delta_{k_3 k} \delta_{k_1 k} \rho_{k'k'}(\varphi_l) 
\delta_{k_4 k'} \delta_{k_2 k'} \nonumber \\   
& = & \frac{1}{2} \sum_{k,k'} {\bar{v}}_{k k'k k'} \cdot 
\sum_l y_l \rho_{kk} (\varphi_l) \rho_{k'k'} (\varphi_l) \nonumber \\
& = & \frac{1}{2} \sum_{k,k' > 0} ({\bar{v}}_{k k'k k'} + 
{\bar{v}}_{\bar{k} k'\bar{k} k'} + {\bar{v}}_{k \bar{k'}k \bar{k'}} + 
{\bar{v}}_{\bar{k}\bar{k'}\bar{k}\bar{k'}} ) 
\sum_l y_l \rho_{kk} (\varphi_l) \rho_{k' k' } (\varphi_l). 
\label{hfenergy}
\end{eqnarray}
Taking into account that 
$y_l (\varphi_l) \approx \langle \Phi| e^{\img \varphi_l \hat{N}} | \Phi \rangle$, 
we see that the possible poles of 
$\rho_{kk}(\varphi_l)$ and $\rho_{k' k'}(\varphi_l)$ are canceled when $k \neq k'$,
i.e. we may only have  problems if $k = k'$. 
The contribution to the  Hartree- Fock energy in this case is given by
\begin{eqnarray}
[E_{hf}^P]_{pole} &=& \sum_{k > 0} \sum_l {\bar{v}}_{k \bar{k} k \bar{k}} \, \, y_{l} 
\frac{v_k^4e^{4 \img \varphi_l}} { {(u_k^2 + v_k^2 e^{2 \img \varphi_l})}^2}
\nonumber \\
&=& {\mathcal N}
\sum_{k > 0} \sum_l {\bar{v}}_{k \bar{k} k \bar{k}}
\frac{v_k^4e^{4 \img \varphi_l}} { {(u_k^2 + v_k^2 e^{2 \img \varphi_l})}}
 \cdot \prod_{m>0,m \ne k} 
(u_m^2 + v_m^2 e^{2 \img \varphi_l} ).
\label{dirterm}
\end{eqnarray}
where we have made use of eq.~(\ref{yltau})
and $ {\mathcal N} = \left[ \sum_{l=1}^L \langle {\Phi} |  
e^ {i \varphi_l \hat{N}} | \Phi \rangle \right ]^{-1}$.
This contribution clearly diverges for $u_k^2=v_k^2$ and $\varphi_l= \pi/2$.
The pairing energy term is given by eq.~(\ref{eq:EPPA}), in the canonical basis it takes
the form
\begin{eqnarray}
E_{pai}^P &  = & \frac{1}{4} \sum_{k,k' > 0} \sum_l y_l  ( 
{\bar{v}}_{\bar{k} k \bar{k}' k' } \kappa_{\bar{k} k}^{01} 
\kappa_{\bar{k}' k '}^{10} +  
{\bar{v}}_{k \bar{k}  \bar{k}' k'} \kappa_{k \bar{k}}^{01} 
\kappa_{\bar{k}' k '}^{10} + \nonumber \\  
& + & {\bar{v}}_{\bar{k} k k' \bar{k}'} \kappa_{\bar{k} k}^{01} 
\kappa_{k' \bar{k}'}^{10} +  
      {\bar{v}}_{k \bar{k} k' \bar{k}'} \kappa_{k \bar{k} }^{01} 
      \kappa_{k' \bar{k}'}^{10}   )  \nonumber \\
& = & \sum_{k k' > 0} \sum_l y_l {\bar{v}}_{k \bar{k} k' \bar{k}'} 
\kappa_{k \bar{k}} ^ {01} \kappa_{k' \bar{k}'}^{10}. 
\label{paienergy}
\end{eqnarray}
As before, $y_l$  cancels one possible pole, and if $k = k'$ the contribution 
to the energy is
\begin{eqnarray}
[E_g^P]_{pole} &=& \sum_{k > 0} \sum_l  {\bar{v}}_{ k \bar{k} k \bar{k}} \,  \, y_{l} 
\frac{u_k^2 v_k^2 e^{2 \img \varphi_l}} { {(u_k^2 + v_k^2 e^{2 \img \varphi_l})}^2}
\nonumber \\
&= & {\mathcal N}\sum_{k > 0} \sum_l {\bar{v}}_{k \bar{k} k \bar{k}}
\frac{u_k^2 v_k^2 e^{2 \img \varphi_l}} { {(u_k^2 + v_k^2 e^{2 \img \varphi_l})}}\cdot
  \prod_{m>0,m \ne k} 
(u_m^2 + v_m^2 e^{2 \img \varphi_l} )
\label{exchterm}
\end{eqnarray}
which also diverges for $u_k^2=v_k^2$ and $\varphi_l= \pi/2$.
 The total contribution to the energy is given by the sum of eq.(\ref{dirterm})
 and eq.(\ref{exchterm}), we obtain
\begin{eqnarray}
[E_{hf}^P]_{pole}+ [E_g^P]_{pole}&=& \sum_{k > 0} \sum_l  {\bar{v}}_{ k \bar{k} k \bar{k}} \,  \,  
\frac{y_{l} v_k^2 e^{2 \img \varphi}} { {(u_k^2 + v_k^2 e^{2 \img \varphi_l})}^2}
\nonumber \\
&= & 
{\mathcal N}\sum_{k > 0} \sum_l {\bar{v}}_{k \bar{k} k \bar{k}}
 v_k^2 e^{2 \img \varphi_l}
\cdot \prod_{m>0,m \ne k} (u_m^2 + v_m^2 e^{2 \img \varphi_l} )
\label{totalterm}
\end{eqnarray} 
 
which clearly does not diverge at $u_k^2=v_k^2$ and $\varphi_l= \pi/2$.

This result shows that in case of particle number projection one cannot
arbitrarily neglect exchange terms of any component of the two-body potential,
including the Coulomb one.

\subsection{Density--dependent interactions.}

In the previous demonstration we have not considered the case of
density--dependent interactions $\bar{v}_{k_1k_2k_3k_4}$ in the analysis of the
convergence. 
We shall distinguish between the two choices for the Hamiltonian density
in order to study this problem.

\subsubsection{Projected density.}
\label{sect:approden}
Taking into account the expression~(\ref{eq:EPDD1}), one sees
that the density--dependent energy
depends only on the proton and neutron projected densities.  We must
therefore analyze the behavior of these densities in the presence of poles.
The projected density is given by 
\begin{eqnarray}
\rho^P_{k_1k_2} & = & \sum_l y_l \rho_{k_1k_2}(\varphi_l). 
\end{eqnarray}
As we saw in the previous section, any pole of 
$\rho_{k_1k_2}(\varphi_l)$ will be canceled by a zero of the same
order from the factor $y_l$.
Therefore, if we take the projected density for the density dependent term, 
the total energy including this term is well--defined.

\subsubsection{Mixed density.}
\label{sect:apmixden}

In this case the energy of the density--dependent term is
given by eq.~(\ref{eq:EPDD2}). As we can see in this expression 
and in eq.~(\ref{eq:gammatil}), 
the densities $\rho (\varphi_{l_{\tau})}$ and the norms $y_{l_{\tau}}$
appear pairwise indicating a cancellation of possible poles and zeros.
However, problems may arise if the interaction ${V}_{DD}$ itself has some poles. 
In this prescription ${V}_{DD}$ depends on  
\( \rho (\vec{r},\varphi_l) \) given by eq.~(\ref{mixdens}).
Writing the density matrix elements in the canonical basis we find that

\begin{equation}
 \rho (\vec{r},\varphi_l )=\sum _{k}\left( f_{kk}\left( \vec{r}\right) 
+f_{\bar{k}\bar{k}}\left( \vec{r}\right) \right) 
\frac{v_{k}^{2}e^{2i\varphi_l }}{u_{k}^{2}+v_{k}^{2}e^{2i\varphi_l }}
\end{equation}

where the functions \( f_{kk'}(\vec{r}) \) are defined in eq.~(\ref{rhodef}).
Clearly, if one of the \( v_{k}^{2} \), let's say \( v_{k_{0}}^{2} \), equals
\( 1/2 \) then the density is singular at \( \varphi_l =\pi /2 \) and in the
neighborhood of this point it behaves as
\begin{equation}
\rho (\vec{r},\varphi_l )\sim -\frac{i}{2}\left( f_{k_{0}k_{0}}
\left( \vec{r}\right) +f_{\overline{k}_{0}\overline{k}_{0}}
\left( \vec{r}\right) \right) \frac{1}{\left( \varphi_l -\pi /2\right) }
\end{equation}
which is clearly divergent. This singularity, however, does not pose
any real drawback as the density dependence of the interaction is proportional
to \( \rho (\vec{r},\varphi_l )^{\alpha } \) with \( \alpha =1/3 \) and 
therefore the singularity is integrable.

%\newpage

%\vskip 3. cm


\begin{thebibliography}{99}
\bibitem{RS.80} P. Ring and P. Shuck, { \it The Nuclear Many Body Problem} 
(Springer-Verlag, Berlin, 1980).
\bibitem{PNP.orig} K. Dietrich, H. J. Mang and J. H. Pradal,
Phys. Rev. {\bf 135} (1964) B22.
\bibitem{EMR.80} J.L. Egido, H.J. Mang, P. Ring,
   Nucl. Phys. A 339 (1980) 390-414.
\bibitem{Egi82}
J. L. Egido and P. Ring,
Nucl. Phys. { A 383}, 189 (1982);
Nucl. Phys. { A 388},  19 (1982).
\bibitem{LN.orig} H.J. Lipkin,, Ann. Phys. (NY) 12 (1960) 425;
Y. Nogami Phys. Rev. B 134 (1964)313
\bibitem{Gall94}
B. Gall, P. Bonche, J. Dobaczeski, H. Flocard, P. H. Heenen,
Z. Phys. {A 348}(1994) 183 .
\bibitem{Bon96}
P. Bonche, H. Flocard, P. H. Heenen, 
Nucl. Phys. { A 598}, 169 (1996).
\bibitem{Val96} A. Valor, J.L. Egido and L. M. Robledo,
Phys. Rev. {C 53} (1996) 172.
\bibitem{Ring.LN}A.V. Afanasjev et al., Nucl. Phys. { A 676}(2000) 196.
\bibitem{Gog80}
J. Decharg\'e and D. Gogny,
Phy. Rev. {C 21}(1980) 1568 .

\bibitem{Fom}V.N. Fomenko. J. Phys. (G.B) A 3 (1970) 8.

\bibitem{Bal69}
R. Balian and E. Brezin, Nuovo Cimento {64} (1969) 37.

\bibitem{VER.00} A. Valor. J. L. Egido and L. M. Robledo, 
Nucl. Phys. {A 665}(2000)46-70.
\bibitem{Bon90}
P. Bonche, J. Dobaczeski, H. Flocard, P. H. Heenen and J. Meyer, {Nucl.
Phys.} {A 510} (1990) 466. 

\bibitem{VER.97} A. Valor, J.L. Egido and L. M. Robledo,
Phys. Lett. {B 392} (1997) 249.

\bibitem{Egi95}
J.L. Egido, J. Lessing, V. Martin and L.M. Robledo, {Nucl. Phys.} 
{A 594} (1995) 70. 

\bibitem{Doe.98} F. Doenau, 
Phy. Rev. {C  58} (1998) 872.

\bibitem{AER.01} M. Anguiano, J.L. Egido and L.M. Robledo,  
Nucl. Phys. { A 683} (2001)227-254. 
\bibitem{GG.83} M. Girod and B. Grammaticos, 
Phys. Rev. {C 27} (1983) 2317.

\bibitem{ER.93}
J. L. Egido and L. M. Robledo,
Phys. Rev. Lett. { 70} (1993) 2876.

\bibitem{Goo79}
A.L. Goodman, in { Advances in Nuclear Physics} {11} (1979) 263. 

\bibitem{Carlo} K. W. Schmid and F. Gruemmer, 
              Rep. Prog. Phys. 50 (1987) 731.
\bibitem{ER164} E.N. Shurshikov and N.V. Timofeeva, 
Nuclear Data Sheets {\bf 65} (1992)365.

\bibitem{CEM.95} E. Caurier, J.L. Egido, G. Martinez-Pinedo, A. Poves, J.
Retamosa, L. M. Robledo and A. Zuker, {\em Phys. Rev. Lett.} {\bf 75}, 2466
(1995)
\bibitem{MPR.96}  G. Martinez-Pinedo, A. Poves,  L. M. Robledo, 
E. Caurier, F. Nowacki, J. Retamosa and A. Zuker, {\em Phys. Rev. } 
{\bf C54}, R2150 (1996)
\bibitem{exp.C48} J.A. Cameron et al., Phys. Rev C49, 1347 (1994); J.A. Cameron et
al., Phys. Lett. B319, 58 (1993); T.W. Burrows, Nucl. Data Sheets 68, 1
(1993).
\bibitem{exp.C50} T.W. Burrows, Nuclear Data Sheets (1990); S.M. Lenzi et al.,
Phys. Rev C56, 1313 (1997).

\bibitem{T=0pair} A. Poves and G. Martinez-Pinedo,
Phys. Lett. {\bf B430} (1998)203.
\bibitem{Go.75} D. Gogny, in ``Nuclear selfconsistent fields", Eds.
G. Ripka and M. Porneuf (North Holland, 1975).
\bibitem{Ber91}
J. F. Berger, M. Girod and D. Gogny,
Comp. Phys. Comm. {\bf 63}, 365 (1991). 

\end{thebibliography}
\end{document}